\documentclass[twocolumn]{aastex631}

\usepackage{textcomp, gensymb}

\usepackage{amsmath}
\usepackage{booktabs}

\begin{document} 
\title{Stellar and orbital characterization of three low mass M dwarf binary stars with dynamical spectroscopy from the Habitable Zone Planet Finder\footnote{Based on observations obtained with the Hobby-Eberly Telescope (HET), which is a joint project of the University of Texas at Austin, the Pennsylvania State University, Ludwig-Maximillians-Universitaet Muenchen, and Georg-August Universitaet Goettingen. The HET is named in honor of its principal benefactors, William P. Hobby and Robert E. Eberly.
}}

\author[0009-0003-2604-3666]{Suhani Surana}
\affiliation{Steward Observatory and Department of Astronomy, University of Arizona, 933 N. Cherry Avenue, Tucson, AZ 85721, USA}
\affiliation{Department of Physics, University of Arizona, 1118 E. 4th Street, Tucson, AZ 85721, USA}
\affiliation{Department of Computer Science, University of Arizona, Tucson, AZ 85721, USA}
\email{suhanisurana@arizona.edu}

\author[0000-0003-4384-7220]{Chad F.\ Bender}
\affiliation{Steward Observatory and Department of Astronomy, University of Arizona, 933 N. Cherry Avenue, Tucson, AZ 85721, USA}
\email{cbender@arizona.edu}

\author[0000-0003-4835-0619]{Caleb I. Ca\~nas}
\altaffiliation{NASA Postdoctoral Program Fellow}
\affiliation{NASA Goddard Space Flight Center, 8800 Greenbelt Road, Greenbelt, MD 20771, USA}

\author[0000-0001-9626-0613]{Daniel M.\ Krolikowski}
\affiliation{Steward Observatory and Department of Astronomy, University of Arizona, 933 N. Cherry Avenue, Tucson, AZ 85721, USA}

\author[0000-0001-9662-3496]{William D. Cochran}
\affiliation{McDonald Observatory and Center for Planetary Systems Habitability, The University of Texas, Austin TX 78712 USA.}

\author[0000-0002-0885-7215]{Mark Everett}
\affiliation{U.S. National Science Foundation, National Optical-Infrared Astronomy Research Laboratory, 950 N. Cherry Ave., Tucson, AZ 85719, USA}

\author[0000-0002-5463-9980]{Arvind F. Gupta}
\affiliation{U.S. National Science Foundation, National Optical-Infrared Astronomy Research Laboratory, 950 N. Cherry Ave., Tucson, AZ 85719, USA}

\author[0000-0001-8401-4300]{Shubham Kanodia}
\affiliation{Earth and Planets Laboratory, Carnegie Science, 5241 Broad Branch Road, NW, Washington, DC 20015, USA}

\author[0000-0001-9596-7983]{Suvrath Mahadevan}
\affiliation{\PSUAA}
\affiliation{\PSUCEHW}  
\newcommand{\PSUAA}{Department of Astronomy \& Astrophysics, 525 Davey Laboratory, 251 Pollock Road, Penn State, University Park, PA, 16802, USA}
\newcommand{\PSUCEHW}{Center for Exoplanets and Habitable Worlds, 525 Davey Laboratory, 251 Pollock Road, Penn State, University Park, PA, 16802, USA}

\author[0000-0002-0048-2586]{Andrew Monson}
\affiliation{Steward Observatory and Department of Astronomy, University of Arizona, 933 N. Cherry Ave., Tucson, AZ 85721, USA}

\author[0000-0001-8720-5612]{Joe P.\ Ninan}
\affiliation{Department of Astronomy and Astrophysics, Tata Institute of Fundamental Research, Homi Bhabha Road, Colaba, Mumbai 400005, India}

\author[0000-0003-1324-0495]{Leonardo A. Paredes}
\affiliation{Steward Observatory and Department of Astronomy, University of Arizona, 933 N. Cherry Avenue, Tucson, AZ 85721, USA}

\author[0000-0003-0149-9678]{Paul Robertson}
\affiliation{Department of Physics \& Astronomy, University of California, Irvine, CA 92697, USA}

\author[0000-0001-8127-5775]{Arpita Roy}
\affiliation{Astrophysics \& Space Institute, Schmidt Sciences, New York, NY 10011, USA}

\author[0000-0002-4046-987X]{Christian Schwab}
\affiliation{School of Mathematical and Physical Sciences, Macquarie University, Balaclava Road, North Ryde, NSW 2109, Australia}

\author[0000-0001-7409-5688]{Gudmundur Stefansson}
\affiliation{Anton Pannekoek Institute for Astronomy, 904 Science Park, University of Amsterdam, Amsterdam, 1098 XH}

\begin{abstract}
 Theoretical models of low-mass stars continue to be discrepant with observations when used to examine the mass–radius relationship and other physical parameters of individual stars. High-resolution spectroscopy that leads to dynamical measurements of binary stars can directly improve these models. 
 We have been using the Habitable-zone Planet Finder spectrograph to monitor binary stars with M dwarf components. Here, we measure the orbital and stellar parameters for three such systems: LSPM J0515+5911, NLTT 43564, and NLTT 45468. Each system has dozens of spectra obtained over a baseline of several years. None of the systems appear to be eclipsing, so our ability to turn them into true benchmark binaries with purely dynamical measurements is limited. We use literature photometry to estimate each system's spectral energy distribution and utilize models in combination with detection limits of our spectroscopic measurements to probe characteristics of the companions.
 LSPM J0515+5911 is a double-lined spectroscopic binary with period of $126.948
\pm 0.029$ days and derived minimum masses, $M_1\sin^3i =0.058 \pm 0.002$ $M_\odot$ and $M_2\sin^3i =0.046 \pm 0.001$ $M_\odot$ for the primary and secondary components, respectively. We solved NLTT 43564 with period of $1877 \pm 24$ days and NLTT 45468 with period of $9.686 \pm 0.001$ days as single lined systems, and modeled the primary masses to be $M_1 = 0.32\pm{0.02}$ $M_\odot$ and $M_1 = 0.35^{+0.02}_{-0.07}$ $M_\odot$, respectively.

\end{abstract}


\section{Introduction} \label{sec:intro}

A large fraction of the star systems in our galaxy contain multiple stellar components \citep{1991A&A...248..485D, 2010ApJS..190....1R}. Binary stars are empirical probes for theoretical models of star formation and evolution, and underpin much of our understanding of stellar astrophysics.
M dwarfs make up the majority of stars in the solar neighborhood \citep{2006AJ....132.2360H, 2018AJ....155..265H}, but have been difficult to observe due to their small size and intrinsically low luminosity. As a result of limited observations and the complexity of their structure and evolution, late M dwarfs \citep[e.g.,][]{2013ApJ...776...87S, 2023Univ....9..498M}, are less well understood than Sun-like (FGK) stars. Directed observations of M dwarfs, when combined with stellar models, can broaden our understanding of these low-mass stellar populations. 
They are also promising targets for exoplanet searches due to their low luminosities, which place the habitable zone closer to the star than around Sun-like stars. This results in shorter orbital periods and more favorable signals (e.g., contrast ratios, semi-amplitude velocities) that are easier to observe than for analogous planets around Sun-like stars. The proximity of the habitable zone facilitates atmospheric characterization, with current or upcoming instruments, such as the James Webb Space Telescope \citep{2021PASP..133e4401G}, ground-based Extremely Large Telescopes \citep{2023AJ....165..267H}, and the future Habitable Worlds Observatory \citep{2024arXiv240208038T}. Characterizing such systems, however, requires well-constrained stellar and orbital parameters \citep[e.g.,][]{2013ApJ...765..131K,2017ApJ...845....5K,2016PhR...663....1S}.  
Model isochrones are not well-constrained for M dwarfs because of the relative dearth of empirical M dwarf masses \citep{2018MNRAS.481.1083P}. 

Double-lined spectroscopic binaries provide dynamical mass measurements that are mostly model independent. Measurements from these star systems help refine and improve models of stellar structure \citep{2023Univ....9..498M}. In this paper, we present precision spectroscopic measurements for three systems:  LSPM 0515$+$5911  whose primary star has a spectral type of M7.0V \citep{2003AJ....125.1598L}; NLTT 43564 (LSPM J1645+6057) whose primary star has a spectral type of M3.5V \citep{2015A&A...577A.128A} and NLTT 45468 (LSPM J1746+6024) whose primary star has a spectral type of M4.5 \citep{2004AJ....128..463R}.  Table~\ref{tab:general} provides a summary of literature values for these systems. 

We started observing these and other M dwarf targets in 2019 with the Habitable-zone Planet Finder (HPF) spectrograph \citep{2012SPIE.8446E..1SM}, as part of a Guaranteed Time Observation program designed to search for exoplanets around M dwarfs. This program has produced numerous planet discoveries and followup confirmations \citep[e.g.,][]{2021yCat..51620061L, 2023Sci...382.1031S, 2025AJ.170.279B}.  We quickly identified these targets as spectroscopic binaries based on large orbital motion in the primary velocities, and because of their low stellar masses we continued observing them at a low cadence over the subsequent years. We used our HPF spectra to measure the radial velocities of individual binary components, derived orbital parameters for each system, and combined these with photometry from the Gaia \citep{2018GAIA, 2023GAIA}, Pan-STARRS \citep{2016arXiv161205560C}, APASS \citep{2015AAS...22533616H}, and 2MASS \citep{2003AJ....125.1598L} surveys to estimate the physical parameters for individual stellar components. We also utilized high contrast and high resolution measurements from the speckle imagers NESSI and 'Alopeke \citep{2018PASP..130e4502S,10.1117/12.2311539}, and imaging from Robo-AO \citep{2012SPIE.8447E..2OR}, to look for potential contamination from background sources.

In \S  \ref{sec:Observations},  we describe the facilities and data products used for this work. In  \S \ref{sec:Data Analysis} we discuss the detailed analysis of the data, including extracting radial velocities (RVs), data analysis and modeling methods, and the determination of stellar parameters. In  \S  \ref{sec:Results} we describe the three binary stars that have been analyzed, provide some discussion of the systems.  In \S\ref{sec:summary} we conclude by summarizing the work discussed and provide remarks about the future of this HPF binary star program.

\begin{deluxetable*}{lcccc}
\tabletypesize{\small}
\tablecaption{Primary Star Parameters.\label{tab:general}}
\tablehead{
\colhead{Parameter} & 
\colhead{LSPM J0515+5911} &
\colhead{NLTT 45468} &
\colhead{NLTT 43564}
}
\startdata
\hline
\multicolumn{4}{l}{\textbf{Primary Star Parameters}} \\
\hline
Gaia DR3 & 281261607033049472 &  1435275923117125760 & 1624385699685942912 \\
TIC ID & 87302025 & 219807052 & 289536553 \\
RA (HH:MM:SS) & 05:15:30.95	 &  17:46:11.64 & 16:45:55.05 \\
Dec (DD:MM:SS) & +59:11:17.44 &  +60:24:26.22 & +60:57:03.96 \\
Proper motion (RA) (mas yr$^{-1}$) & $116.467 \pm 0.211$  & $-79.414 \pm 0.021$ & $22.24 \pm 0.30$ \\
Proper motion (DEC) (mas yr$^{-1}$) & $-1005.994 \pm 0.182$  & $214.516 \pm 0.0170$ & $-279.60 \pm 0.30$ \\
Parallax (mas) & $63.36 \pm 0.24$  & $	35.57 \pm 0.01$ & $46.68 \pm 0.09$ \\
RUWE & 4.631  & 1.235 & 17.164 \\
\hline
\multicolumn{4}{l}{\textbf{Photometry}} \\
\hline
Pan-STARRS $g^{~b}$ &  $18.687 \pm 0.016 ^{a}$  & $15.465 \pm 0.021$ & \nodata \\
Pan-STARRS $r^{~b}$ &  $17.311 \pm 0.012 ^{a}$  & $14.199 \pm 0.005$ & \nodata  \\
Pan-STARRS $i ^ {~b}$ &  $14.661 \pm 0.002^{a} $  & $12.730 \pm 0.120$ & \nodata \\
Pan-STARRS $z ^ {~b}$ &  $13.593 \pm 0.017^{a}$ & \nodata & \nodata \\
Pan-STARRS $y ^ {~b}$ &  $12.868 \pm 0.004^{a}$  & $11.504 \pm 0.045$ & \nodata\\
Gaia $BP ^ {~c}$       &  $18.335 \pm 0.013^{a}$  & $15.0819 \pm 0.0037$ & $13.7076\pm0.0030^{a} $  \\
Gaia $RP ^ {~c}$       &  $13.8142 \pm 0.0011^{a}$ & $12.0067 \pm 0.0013$ & $11.0165\pm 0.0013^{a}$ \\
Gaia $G ^ {~c}$       &  $15.2895 \pm 0.0007^{a}$  & $13.2902 \pm 0.0005$ & $12.2114\pm0.0008^{a}$ \\
APASS $B  ^ {~d}$       &  \nodata  & $16.638 \pm 0.204^{a}$ & $14.973 \pm 0.124$ \\
APASS $V ^ {~d}$       &  \nodata  & $14.702	\pm 0.067^{a}$ & $13.411 \pm 0.036$ \\
APASS $g ^ {~d}$       &  \nodata  & $15.608\pm 0.170^{a}$ & $14.204 \pm 0.049$ \\
APASS $r ^ {~d}$       &  \nodata  & $14.106\pm 0.027^{a}$ & $12.855 \pm 0.040$ \\
APASS $i ^ {~d}$       &  \nodata  & $12.298 \pm	0.164^{a}$ & $11.382 \pm 0.032$ \\
2MASS $J ^ {~e}$       &  $11.320\pm	0.025$  &  $10.011	\pm 0.027^{a}$   & $9.393 \pm 0.026$  \\
2MASS $H ^ {~e}$       &  $10.661\pm	0.019$  & $9.372	\pm 0.024^{a}$ & $8.840 \pm 0.027$ \\
2MASS $K_s ^ {~e}$     &  $10.319\pm	0.018$ & $9.039	 \pm 0.018^{a}$ & $8.568 \pm 0.025$  \\
WISE $1 ^ {~f}$     &  $10.001 \pm 0.023$
 & \nodata & \nodata  \\
WISE $2 ^ {~f}$     &  $9.792 \pm 0.020$ & \nodata & \nodata  \\
WISE $3 ^ {~f}$     &  $9.440 \pm 0.040$ & \nodata & \nodata  \\
WISE $4 ^ {~f}$     &  $9.023 \pm 0.464$ & \nodata & \nodata  \\
\enddata

\tablenotetext{a}{Not used in the SED Fit described in \S\ref{sec:SED}}
\tablerefs{
$^{b}$ Pan-STARRS: \citep{2016arXiv161205560C};
$^{c}$ Gaia DR2: \citep{2018GAIA}, Gaia DR3: \citep{2023GAIA};
$^{d}$ APASS: \citep{2015AAS...22533616H};
$^{e}$ 2MASS: \citep{2003yCat.2246....0C}; $^{f}$ WISE: \citep{2014yCat.2328....0C}
}

\end{deluxetable*}

\section{Observations} \label{sec:Observations}

\subsection{High-resolution spectroscopy with HPF}

The spectroscopic observations listed in Table \ref{tab:rv_single} were all obtained with HPF  \citep{2012SPIE.8446E..1SM}, on the 10~m Hobby Eberly Telescope (HET) \citep{1998SPIE.3352...34R, 2005AIPC..752....3R, 2021AJ....162..298H}, at McDonald Observatory, Texas. HPF is a high-resolution $(R \sim 53,000)$, 1.7\arcsec{} fiber-fed \citep{2018SPIE10702E..6QK}, temperature controlled \citep{2016SPIE.9908E..71S}, near-infrared ($\lambda\sim 8,080-12,780$~\AA) spectrograph. HPF uses gold-coated mirrors, a mosaic echelle grating, and a single Teledyne Hawaii-2RG (H2RG) NIR detector with a 1.7-micron cutoff covering parts of the z, Y and J NIR bands \citep{2014SPIE.9147E..1GM}. 
HPF observations are queued and executed by the resident astronomers at the HET \citep{2007PASP..119..556S}. This allows us to target observations of binary stars at specific orbital phases and facilitates long-term observing programs needing measurements on monthly or every few months cadence.   

We executed observations at strategic orbital phases to cover the full binary orbits of our target systems with a signal-to-noise ratio (S/N) sufficient to probe for a signal from the secondary star. HPF operations impose a maximum single exposure time of 945 seconds. As this program progressed, we adapted the requested exposure time of each target to adjust the S/N, and to stay within time allocated for the program. In some cases, Table~\ref{tab:rv_single} lists back-to-back exposures on a single night; this represents exposure sequences structured to keep under the maximum individual exposure time, and which we subsequently analyzed independently. In May 2022, HPF required maintenance that included a vacuum cycle, which resulted in two RV eras, the ``pre" and ``post", corresponding to RVs before and after this cycle. There is a small offset in the RVs at the m~s$^{-1}$ level observed across these eras.  In the analysis in \S3, we treat RVs from these eras as coming from two different instruments. 

All spectra were extracted using the standard HPF pipeline which uses the \texttt{HxRGproc} code\footnote{\url{https://github.com/indiajoe/HxRGproc}} which processes the data and performs bias noise removal, nonlinearity correction, cosmic-ray correction, and slope/flux and variance image calculation \citep{2019JATIS...5d1511N}. Observations in the pre era were extracted using a classical optical extraction, which automatically removes the signature of the echelle blaze. In the post era, the extraction changed to a flat-relative algorithm, and the blaze was manually removed based on a static correction obtained by sum extracting very high S/N continuum lamp spectra and smoothing them to remove detector pixel QE variations. This pipeline change was implemented because of a software upgrade to the computer cluster that runs the pipeline, which made utilizing the original code difficult. The updated pipeline produces functionally identical extracted spectra to the old pipeline.

Standard HPF operations obtain Laser-Frequency Comb (LFC) wavelength calibration observations throughout the night \citep{2019Optic...6..233M}, and the pipeline interpolates these to derive the wavelength solution at the time of any specific science observations \citep{2020AJ....159..100S}.
\S\ref{subsubsec: Telluric Correction} describes our procedure for correcting telluric absorption. We computed barycentric corrections using \texttt{barycorrpy}\footnote{\url{https://github.com/shbhuk/barycorrpy}} \citep{2018ascl.soft08001K} based on algorithms presented in \citet{Wright_2014}.

LSPM 0515$+$5911 was observed on 24 epochs from 2020 January to 2022 February. NLTT 43564 was observed on 73 epochs from 2019 April to 2025 April. NLTT 45468 was observed on 80 epochs from 2019 May to 2025 June. 
The median per-pixel S/N in HPF order index 18 ($\lambda\sim 10,700$~\AA, approximately the center of HPF's bandpass, where the HPF exposure time calculator works), RVs of all HPF observations, and $1\sigma$ uncertainties are reported in Table \ref{tab:rv_single}. For NLTT 45468, the median per-pixel S/N is 39.6 in HPF order index 1  $(8079\text{--}8184\,\text{\AA})$
 and 91.8 in the HPF order index 28 $(12625\text{--}12786\,\text{\AA})$.

\begin{deluxetable*}{ccccccc}
\tablecaption{HPF Spectroscopy Epochs and Radial Velocity Measurements \label{tab:rv_single}}
\tablewidth{0pt}
\tablehead{
\colhead{UT Date} &
\colhead{$\mathrm{BJD_{TDB}}$-2,400,000} &
\colhead{$V_A$} &
\colhead{$\sigma_A$} &
\colhead{$V_B$} &
\colhead{$\sigma_B$} &
\colhead{S/N$^{a}$} \\
\colhead{} &
\colhead{} &
\colhead{(km s$^{-1}$)} &
\colhead{(km s$^{-1}$)} &
\colhead{(km s$^{-1}$)} &
\colhead{(km s$^{-1}$)} &
\colhead{}
}
\startdata
\multicolumn{5}{l}{\textbf{LSPM\,J0515 $+$ 5911 -- 24 epochs}} \\
\hline
2020 Jan 10 & 58858.631101 & -49.220 & 0.061 & \nodata & \nodata & 39.7 \\
2020 Jan 10 & 58858.642743 & -49.164 & 0.062 & \nodata & \nodata & 38.5 \\
2020 Jan 31 & 58879.730610 & -45.966 & 0.063 & -57.056 & 0.587 & 64.2 \\
2020 Jan 31 & 58879.742138 & -45.984 & 0.062 & -56.942 & 0.607 & 62.7 \\
2020 Feb 04 & 58883.699353 & -45.935 & 0.064 & -57.208 & 0.592 & 42.5 \\
2020 Feb 04 & 58883.710931 & -45.914 & 0.064 & -56.558 & 0.767 & 50.5 \\
2020 Sep 02 & 59094.956441 & -67.219 & 0.058 & -29.036 & 0.478 & 40.9 \\
2020 Sep 30 & 59122.888262 & -46.593 & 0.059 & -55.817 & 0.947 & 55.7 \\
2020 Nov 05 & 59158.958795 & -46.699 & 0.057 & -55.220 & 1.326 & 70.4 \\
2020 Nov 05 & 59158.970129 & -46.674 & 0.058 & -55.783 & 0.665 & 71.1 \\
\hline
\multicolumn{5}{l}{\textbf{NLTT 43564 -- 73 epochs}} \\
\hline
2019 Apr 19 & 58592.839000 & -32.872 & 0.061 & \nodata & \nodata & 160.8 \\
2019 Apr 19 & 58592.850000 & -32.844 & 0.061 & \nodata & \nodata & 181.4 \\
2020 Feb 17 & 58896.997000 & -32.633 & 0.061 & \nodata & \nodata & 121.4 \\
2020 Feb 17 & 58897.010000 & -32.637 & 0.067 & \nodata & \nodata & 70.0 \\
2020 Mar 11 & 58919.934000 & -32.623 & 0.058 & \nodata & \nodata & 132.2 \\
2020 Mar 11 & 58919.946000 & -32.646 & 0.059 & \nodata & \nodata & 132.3 \\
2020 Apr 14 & 58953.849000 & -32.573 & 0.062 & \nodata & \nodata & 92.9 \\
2020 Apr 14 & 58953.860000 & -32.633 & 0.061 & \nodata & \nodata & 102.9 \\
2020 Apr 21 & 58960.971000 & -32.605 & 0.056 & \nodata & \nodata & 195.0 \\
2020 Apr 21 & 58960.982000 & -32.594 & 0.055 & \nodata & \nodata & 202.2 \\
\hline
\multicolumn{5}{l}{\textbf{NLTT 45468 -- 80 epochs}} \\
\hline
2019 May 24 & 58627.774000 & -12.083 & 0.070 & \nodata & \nodata & 107.5 \\
2019 May 24 & 58627.786000 & -12.115 & 0.069 & \nodata & \nodata & 106.1 \\
2019 Sep 04 & 58730.665000 & -12.061 & 0.067 & \nodata & \nodata & 109.4 \\
2019 Sep 04 & 58730.677000 & -12.069 & 0.065 & \nodata & \nodata & 115.9 \\
2019 Sep 16 & 58742.594000 & -12.100 & 0.079 & \nodata & \nodata & 59.4 \\
2019 Sep 16 & 58742.607000 & -12.185 & 0.085 & \nodata & \nodata & 53.7 \\
2019 Sep 18 & 58744.633000 & -12.016 & 0.069 & \nodata & \nodata & 93.3 \\
2019 Sep 18 & 58744.645000 & -12.030 & 0.070 & \nodata & \nodata & 89.1 \\
2020 Feb 27 & 58906.996000 & -12.239 & 0.083 & \nodata & \nodata & 56.6 \\
2020 Feb 27 & 58907.008000 & -12.234 & 0.079 & \nodata & \nodata & 62.2 \\
\enddata
\tablecomments{This table is published in its entirety in the machine-readable format. A portion is shown here for guidance regarding its form and content.}
\tablenotetext{a}{{The S/N is calculated as the median of $\mathrm{flux}/\sqrt{\mathrm{variance}}$ in order index~18, near $10,700$~\AA, which is the peak of the HPF throughput.
}}
\end{deluxetable*}

\subsubsection{Telluric Correction}
\label{subsubsec: Telluric Correction}

The Earth's atmosphere absorbs broad swaths of light across the near-infrared bandpass of HPF. To correct this telluric contamination, we apply a method that combines a suite of model atmospheric transmission spectra with a model of the HPF instrumental line spread function (LSF). This paper is the first to use the HPF data processed with this telluric correction algorithm. An accurate LSF model is required to convolve with the transmission model spectra to reflect the telluric contamination as observed in HPF spectra. Below we provide a brief description of our procedure and defer a more detailed discussion of the HPF LSF model and telluric correction to a future paper focused on the method (Krolikowski et al. in prep\footnote{An in-progress version of the code to generate the telluric model can be found at \url{https://github.com/dkrolikowski/eprv_telluric_correction}}).

To prepare for the telluric correction, we first fit a model of the HPF instrumental LSF to LFC spectra \citep{2019Optic...6..233M}. The LFC provides emission lines across the entire HPF bandpass with extremely narrow intrinsic width, meaning that the observed LFC line profile is nearly the instrumental LSF. The model is parameterized as a Gaussian convolved with a trapezoid multiplied by an orthogonal polynomial, with the latter two elements accounting for non-Gaussianity and asymmetry. The resulting LSF model can be evaluated at any given point in an HPF spectrum. We then generate a grid of model transmission spectra for species that absorb within the HPF bandpass (H$_2$O, O$_2$, CO$_2$, and CH$_4$) over a range of water vapor column and zenith angle values. We generate the transmission spectra using line-by-line radiative transfer modeling \citep[LNFLv3.2 and LBLRTM12.15.1,][]{clough1992,clough2005} and the molecular line lists from AER and HITRAN \citep{aer_v3.8.1,hitran2020}. We use the default O$_2$ LBLRTM abundance, fix the CO$_2$ abundance to 420 ppm, and fix the CH$_4$ abundance to 1860 ppb (roughly the average atmospheric concentrations during the observation epochs presented here). Most of the telluric absorption in HPF band is H$_2$O and O$_2$, so CO$_2$ and CH$_4$ are not critical.

To generate a telluric model for each observation, we first interpolate the model transmission spectra grid to the observation's zenith angle. We then convolve the model grid using the HPF LSF model and fit for the water vapor column using select wavelength regions where we expect minimal contamination from stellar spectral lines. Lastly, we combine the model transmission spectra for each molecular species to generate a full telluric line absorption model across the entire HPF spectrum. We corrected for telluric lines in the target spectrum by dividing the science flux by the telluric line absorption model, which although not strictly mathematically correct, is accurate to within a few percent which although not strictly mathematically correct, is accurate to a few tenths of a percent for most lines and better than 2\% for the deepest lines \citep[eg.,][Fig 5]{2022AJ....164..212L}.

\subsection{High-contrast imaging with ground based instruments}
\label{subsec:SpeckleImaging}
We used high-contrast imaging to search for any bright stars near our targets that could contaminate spectra obtained with HPF's 1.7\arcsec{} fiber. Common false positive scenarios that might affect our analysis include blended and/or unresolved background binaries.
where the light from additional stars within the aperture contaminates the signal in our spectrograph;  \citet{2015AJ....149..143F} describe the full suite of contamination possibilities. We observed each of our targets with the NN-Explore Exoplanet Stellar Speckle Imager \citep[NESSI;][]{2018PASP..130e4502S} on the 3.5 m WIYN Telescope at the Kitt Peak National Observatory \citep{2018PASP..130e4502S}. We also inspected archival data from 'Alopeke speckle imaging \citep{2018PASP..130e4502S, 2011AJ....142...19H} and Robo-AO adaptive optics \citep{2012SPIE.8447E..2OR, 2020AJ....159..139L} imaging for NLTT 43464.

\begin{figure}[!t]
\centering
\gridline{\fig{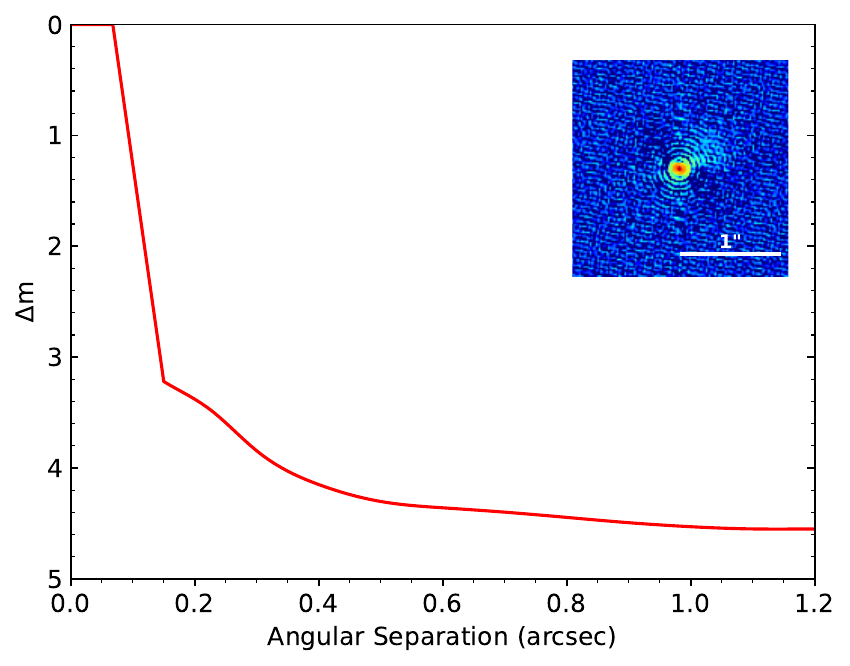}{0.85\columnwidth}{(a) LSPM 0515$+$5911}}
\vspace{-1em}  
\gridline{\fig{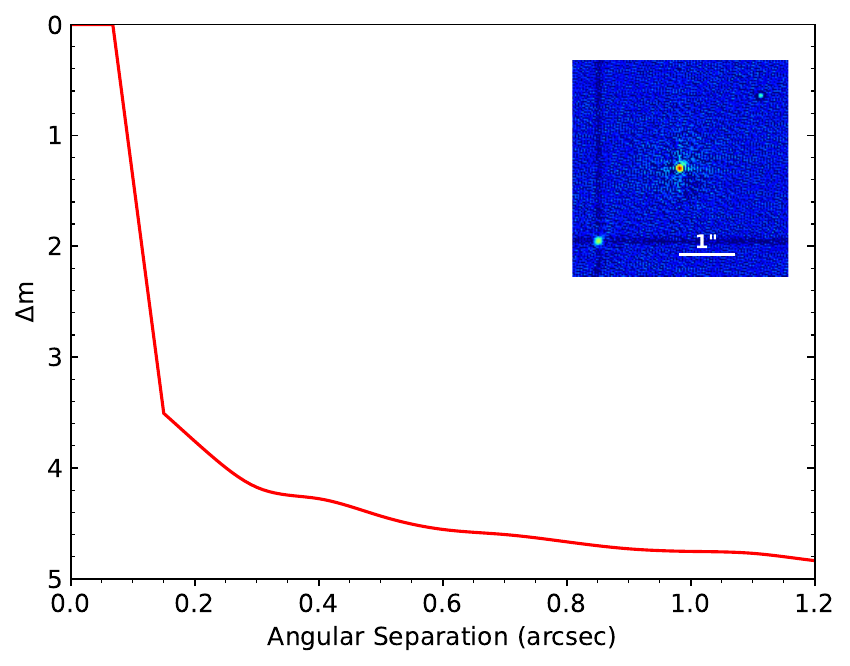}{0.85\columnwidth}{(b) NLTT 45468}}
\vspace{-1em}
\gridline{\fig{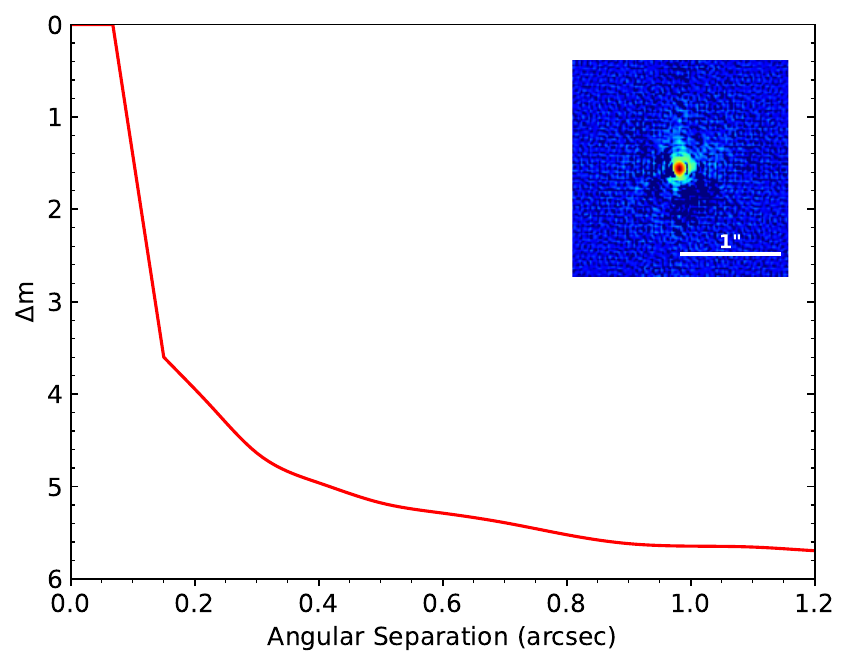}{0.85\columnwidth}{(c) NLTT 43564}}
\vspace{-1em}
\caption{Contrast curves and reconstructed images from the NESSI speckle imager for \emph{a}: LSPM 0515$+$5911, \emph{b}: NLTT 45468, and \emph{c}: NLTT 43564. Neither LSPM 0515$+$5911 nor NLTT 43564 show companions within the field that could contaminate the HPF observations.  NLTT 45468 shows a nearby star at 2.1" in the lower left of the inset in \emph{b}; the signal in the upper right is an artifact of the speckle processing.}
\label{fig:threeplots}
\end{figure}

\subsubsection{NESSI Speckle Imaging}

We obtained speckle observations of LSPM 0515$+$5911 on 2022 April 21, NLTT 43564 on 2022 April 17, and NLTT 45468 on 2022 April 17 using NESSI (PI: Gupta, 2022A-665981). We processed the data using the reduction procedures described in \cite{2011AJ....142...19H}. A 9 minute  sequence with 40 ms exposure times was obtained using NESSI’s red camera, capturing diffraction limited speckle images in the Sloan $z'$ filter. Figure \ref{fig:threeplots}a, \ref{fig:threeplots}b, and \ref{fig:threeplots}c show the contrast curves for NLTT 43564, NLTT 45468 and LSPM J0515 $+$ 5911, respectively.  For NLTT 43564, there are no nearby sources brighter than $\Delta z' = 4.45$ at separations greater than $0\farcs5$. For NLTT 45468, a companion is visible at $\approx$ $2\farcs056$ with delta mag of 1.91 in SDSS $z'$ filter, that for HPF likely results in blending and causes observed distortions in the differential line width (see \S\ref{sec:nltt45468}). This companions was identified by Gaia (Gaia DR3 1435275918821040000) and we discuss it further in \S\ref{sec:nltt45468}.  For LSPM J0515 $+$ 5911, there is no evidence of bright companions or significant flux contamination brighter than $\Delta z' = 4.45$ between 0.2" -- 1.2" from the target.

\subsubsection{Adaptive optics imaging with Robo-AO}

\begin{figure}[!t]
    \includegraphics[width=\linewidth]{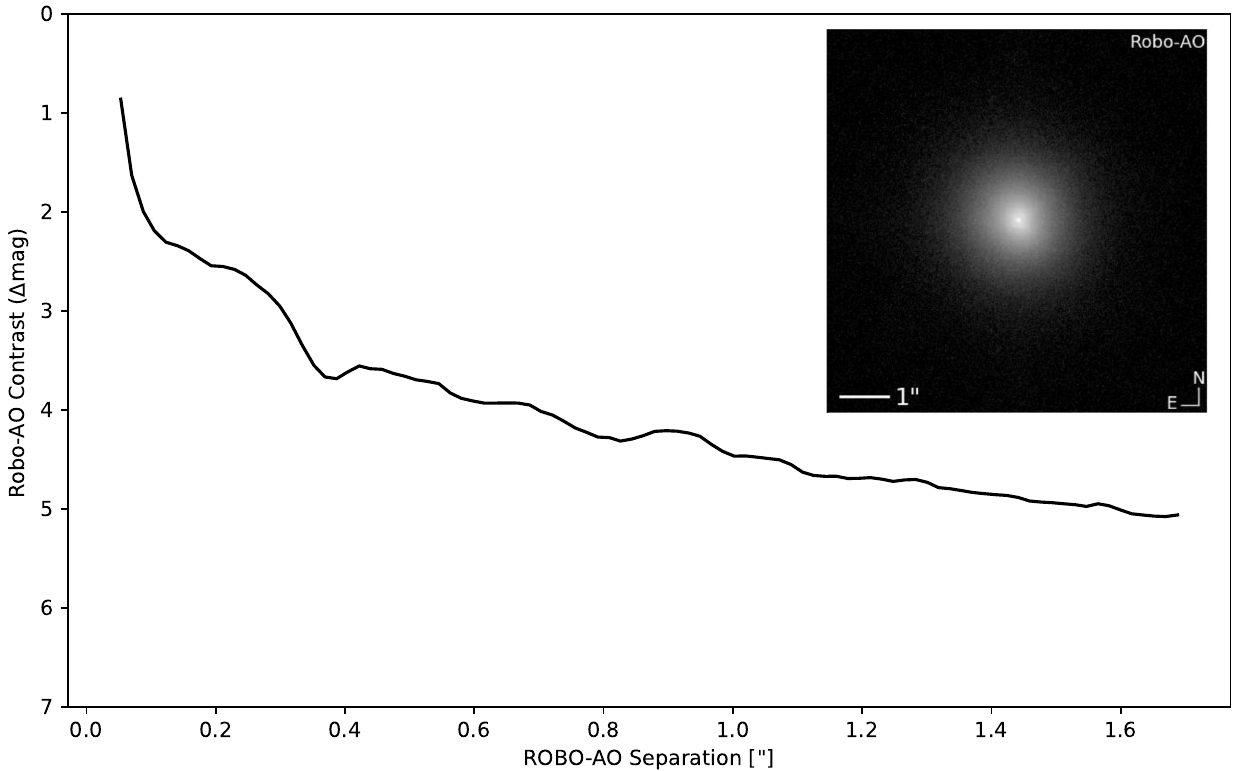}
    \caption{Robo-AO contrast curves for NLTT 43564 obtained with the Robo-AO laser Adaptive Optics system. The curves show the 5$\sigma$ detection limits ($\Delta m$) as a function of angular separation. The inset of 1" $\times$ 1" shows the reconstructed speckle image at 832~nm. No nearby companions are detected within 1.6" angular separation of the target.}
    \label{fig:ROBOAO}
\end{figure}

The Robo-AO laser Adaptive Optics system on the 2.1 m telescope at Kitt Peak National observatory observed NLTT 43564 on 17th April 2017 as part of a survey of M dwarf multiplicity \citep{2020AJ....159..139L}. The data are available through the Exoplanet Follow-up Observation Program \citep{Christiansen_2025, https://doi.org/10.26134/exofop5}. The data were reduced with the Robo-AO pipeline \citep{2014ApJ...791...35L}, which performs PSF subtraction, searches for stellar companions, and computes 5$\sigma$ contrast curves to estimate detection limits.  \cite{2018AJ....155..161Z} presents a detailed description of the data reduction pipeline. Figure~\ref{fig:ROBOAO} shows the Robo-AO contrast curve for NLTT 43564: there are no close companions within 1" of separation. 

\subsubsection{'Alopeke speckle imaging}
\begin{figure}[!t]
    \centering
    \includegraphics[width=1.0\linewidth]{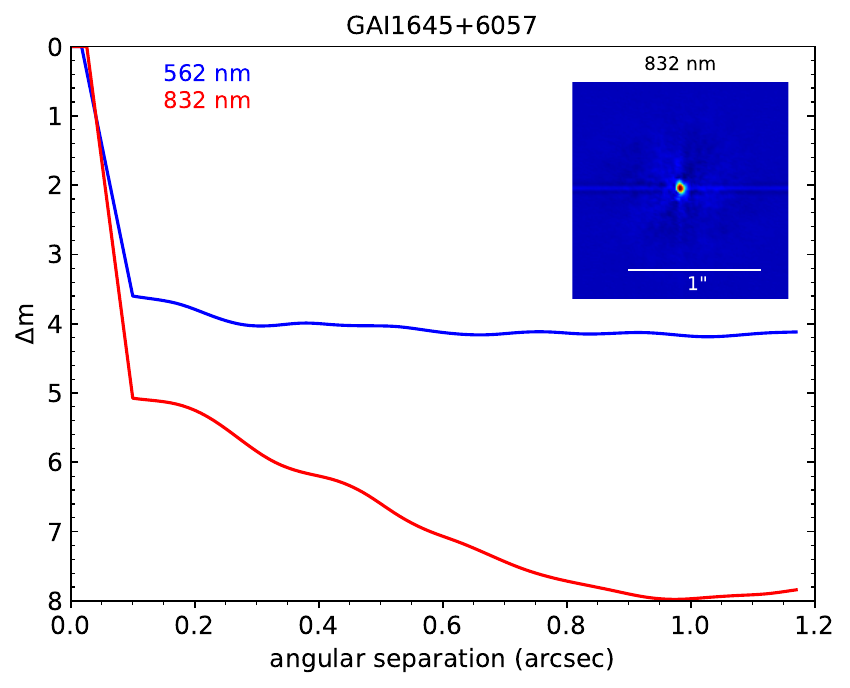}
    \caption{Speckle imaging contrast curves for NLTT 43564 (GAI~1645+6057) obtained with the 'Alopeke Speckle Imaging at Gemini at 562~nm (blue) and 832~nm (red). The curves show the 5$\sigma$ detection limits ($\Delta m$) as a function of angular separation. The inset of 1" $\times$ 1" shows the reconstructed speckle image at 832~nm. No nearby companions are detected within 1.2" angular separation of the target.}

    \label{fig:Gemini}
\end{figure}

NLTT 43564 was observed by the `Alopeke Speckle Imager on the Gemini North telescope on 24th June 2021 (GN-2021A-Q-312, PI: Jao, \cite{, https://doi.org/10.26134/exofop5}) to search for any nearby close companions \citep{2018PASP..130e4502S, 2011AJ....142...19H}. `Alopeke has simultaneous dual channel imaging at 562~nm ($\Delta\lambda = 54$~nm) and 832~nm ($\Delta\lambda = 40$~nm). It has a pixel scale of 0\farcs01~pixel$^{-1}$ and a diffraction limited PSF of $\approx$ 0\farcs02. Figure \ref{fig:Gemini}  shows the contrast curve along with the inset at 832 nm. No bright companions are evident within 1\farcs2, with detection limits of $\Delta m \approx 3.5$ at 0\farcs.18 and $\Delta m \approx 7.5$ at 1\arcsec\ in the 832~nm band, consistent with the single point source.

\vspace{5em}

\section{Data Analysis}
\label{sec:Data Analysis}

\subsection{Analysis of Single Line and Double Line Spectroscopic Binaries}
We used the TODCOR algorithm \citep{1992ASPC...32..164M} to measure the radial velocities of the individual stars in our HPF spectra. TODCOR performs a two-dimensional (2D) cross correlation between the observed spectrum and two theoretical or observed templates shifted by different velocities.
TODCOR is an efficient algorithm that uses the Fast Fourier Transform to compute cross correlations in multiple dimensions, solving for $v_1$ and $v_2$, the radial velocities for the primary and secondary star, and $\alpha$, the flux ratio between the two stars. TODCOR computes the two-dimensional correlation function over a grid of velocity pairs with uniform spacing in velocity, and identifies the combination of $v_1$, $v_2$, and $\alpha$ that maximizes the correlation. This approach can, given sufficient S/N in the observations, disentangle blended spectral features at velocities near the systemic velocity, that would otherwise introduce a bias in a one-dimensional (1D) correlation. 

We estimated the uncertainties of the primary and secondary velocity measurements as a combination of the intrinsic velocity uncertainty of the templates and the uncertainty in measuring the center of the correlation peak due to its intrinsic width, as measured by fitting a quadratic polynomial fit to the correlation peak \citep[see][for more details]{2008ApJ...689..416B}.
We used an IDL based version of TODCOR, \texttt{SXCORR} \citep{2008ApJ...689..416B,2012ApJ...751L..31B}, which implements a convenient interface to interactively explore the correlation space in complicated systems, allows for both one- and two-dimensional cross-correlations, and has an option to lock $\alpha$ to a known fixed value.

Our analysis started with a standard 1D cross correlation to quickly estimate the   velocities of the primary star, and place bounds on the stellar parameters of the template and the plausible velocity parameter space of the secondary star.  We then use those bounds to conduct a more focused 2D analysis with the full TODCOR algorithm.
The stellar parameters of the template, such as spectral type, temperature, surface gravity, metallicity, and rotational velocity, are indicative of the underlying spectral type of the star. However, slight mismatches in the template parameters, have only a minimal impact on the measured RVs in a 1D correlation, and mostly manifest as a change in correlation amplitude. Therefore, we use the maximum correlation value as an indicator of the best matching template. We use the maximum-likelihood method presented in \cite{2003MNRAS.342.1291Z} to combine multiple contiguous regions of spectra, which are separated by order gaps and by areas of uncorrectable telluric absorption. This algorithm is unable to correctly account for changes in  $\alpha$ with wavelength that naturally occur with spectra spanning a wide bandpass, so in that event we follow the procedure outlined in \cite{2012ApJ...751L..31B} to combine spectral regions with varying $\alpha$.

For the SB2 target LSPMJ0515 $+$ 5911, we measured a range of $\alpha$ for the epochs with well-separated peaks in the cross-correlation function, and then locked it for the epochs where the peaks are blended. We used ten spectra with cleanly separated peaks to constrain alpha from 0.17 to 0.20, and then locked $\alpha$ to 0.18, as was done in \cite{2012ApJ...751L..31B}; variations within this range result in negligible variation in the measured RVs. Mismatches between the stellar parameter of the template and observed spectra can significantly impact the estimate of $\alpha$. But for cool M dwarfs, where spectra at HPF's resolution are comprised of numerous blended lines, including molecular features, the measured RVs are stable over small variations in template parameters, while the correlation amplitude can be highly variable. In binaries with low $\alpha$, where the primary star’s light is dominant (i.e., $\alpha$ of a few percent), the RVs are tolerant to changes in $\alpha$ of a factor of two to three. 
For each system, we initially characterized the orbital parameters by fitting the RVs in Table \ref{tab:rv_single} using a Levenberg-Marquardt optimization algorithm to solve Kepler's 2nd law. This approach constrained $\omega$ and $e$, and allowed us to derive initial values for the orbital parameters by minimizing the chi-squared metric. This optimization was then used to inform the priors used in a MCMC optimization for each system, which are listed in Table \ref{table:Stellar}.
The \texttt{exoplanet}\footnote{\url{https://github.com/exoplanet-dev/exoplanet}} \citep{2021JOSS....6.3285F} modeling toolkit was developed for the analysis of exoplanets, but since the underlying orbital physics is the same, we used it for spectroscopic binaries.  \texttt{exoplanet} implements a Hamiltonian Monte Carlo (HMC) parameter estimation from PyMC3 \citep{2015arXiv150708050S} using the Gelman-Rubin statistic of $\hat{R}<1.1$ \citep{2006ApJ...642..505F} to check for convergence.

In \texttt{exoplanet}, we create a probabilistic model in PyMC3 to fit the RV curves, which consist of 8 independent parameters for an SB1 and 9 for an SB2, including the orbital period $P$, the time of inferior conjunction $T_0$, $\sqrt{e} \cos \omega$, $\sqrt{e} \sin \omega$ where $e$ is eccentricity $e$ and $\omega$ is the argument of the periapsis of the star's orbit, semi-amplitude $K_1$, $K_2$ (for SB2s only), a jitter term that includes stellar and instrument RV jitter, $\sigma_{HPF}$, and the systemic velocity, $\gamma$. 
 
After providing initial priors for the six orbital parameters, we modeled the RVs using a standard Keplerian model. To account for pre/post era RV measurements we used two different radial velocity jitters, $\sigma_{HPF_{pre}}$ and $\sigma_{HPF_{post}}$, and two offset parameters $\gamma_{pre}$ and $\gamma_{post}$, that are treated independently. The jitter term is added in quadrature to the observed RV uncertainties when calculating the likelihood function.
The derived parameters, along with their uncertainties, are given in Table \ref{table:Stellar}.

\subsection{Determination of Stellar Parameters}
\label{sec:SED}
To estimate the stellar parameters of the primary star, we utilized \texttt{HPF-SpecMatch}\footnote{\url{https://github.com/gummiks/hpfspecmatch}} \citep{2020AJ....159..100S} which is broadly based on the algorithms described in \cite{2017ApJ...836...77Y}. 
In \texttt{HPF-SpecMatch}, the spectrum is deblazed, sky corrected, barycentric corrected and then is interpolated to a grid of wavelengths spanning $\lambda\sim 8656 - 8768 \mathrm{\AA}$ due to minimal telluric contamination in that region. Then, for each star, the spectrum is rotationally broadened by a projected rotational velocity and multiplied by a Chebyshev polynomial to remove any low-order residual variations in the corrected spectra. \texttt{HPF-SpecMatch} first identifies the best-fit between the target spectrum and each library spectrum and then provides the five best matched templates. It then linearly combines the five spectra into a single composite spectrum and derives stellar parameters $T_{eff}, [Fe/H], \log \textit{g} $ and $v \sin i$ for the primary star as the weighted average of the five best match library stars. To obtain the spectroscopic parameters for SB1s, we ran \texttt{HPF-SpecMatch} on the highest S/N visit for each system, while for the SB2, we chose the epoch for a point at RV gamma, when the separation between the primary/secondary is $\sim$ 0 km/s.  The results are given in Table \ref{table:Stellar}. 
We have intentionally excluded $v \sin i$, because we believe the significance of the values derived for it by \texttt{HPF-SpecMatch} to be low.
To derive the model dependent stellar parameters for the primary star, we used \texttt{astroARIADNE}\footnote{\url{https://github.com/jvines/astroARIADNE}} \citep[henceforth, \texttt{Ariadne};][]{2022MNRAS.tmp..920V} software which is a spectral energy distribution (SED) Bayesian model averaging fitter.  
\texttt{Ariadne} models the available broadband photometry, which is provided in Table \ref{tab:general}. The SED fit adopts Gaussian priors on (i) broadband photometry listed in Table \ref{tab:general}; (ii) the spectroscopic constraints on $\log{g}$, $T_\mathrm{eff}$, and $[Fe/H]$ obtained from our \texttt{HPF-SpecMatch} analysis; (iii) Gaia DR3 parallax and (iv) Bailer-Jones distance \citep{2021AJ....161..147B}. 
\texttt{Ariadne} also applies an upper bound on the visual extinction using the SFD dust maps \citep[defined in][]{1998ApJ...500..525S} from \citet{2011ApJ...737..103S}, evaluated at the Bailer-Jones distance and converted to AV with the \citet{1999PASP..111...63F} reddening law assuming Rv = 3.1. 
Table \ref{tab:general} summarizes the full set of literature photometry used in the joint analysis, and Table \ref{table:Stellar} lists the stellar parameters inferred from \texttt{HPF-SpecMatch} along with their uncertainties.

\section{Results and Discussion}

\label{sec:Results}

\begin{deluxetable*}{cccc}
\tabletypesize{\small}
\tablecaption{Model Priors and Derived Stellar Parameters for the Spectroscopic Binaries\label{tab:priors_results}}
\tablehead{
\colhead{Parameter} &
\colhead{LSPM J0515+5911} &
\colhead{NLTT 45468} &
\colhead{NLTT 43564}
}
\startdata
\hline
\multicolumn{4}{l}{\textbf{Priors}} \\
\hline
\multicolumn{4}{l}{\textbf{Orbital Parameters:}} \\
$P$ (days) & $\mathcal{N}(126.91,1)$ & $\mathcal{N}(9.69,1)$ & $\mathcal{N}(1888.04,25)$ \\
$T_0$ (BJD$_{\text{TDB}}$) & $\mathcal{N}(2459080.8,1)$ & $\mathcal{N}(2458739.5,1)$ & $\mathcal{N}(2459662.6,1)$ \\
$K_1$ (km s$^{-1}$) & $\mathcal{U}(0,15)$ & $\mathcal{U}(0.1,0.3)$ & $\mathcal{U}(0,1)$ \\
$\log(K_2)$ (km s$^{-1}$) & $\mathcal{U}(0,2.708)$ & \nodata & \nodata \\
$\sqrt{e} \cos \omega$ & $\mathcal{U}(-1,1)$ & $\mathcal{U}(-1,1)$ & $\mathcal{U}(-1,1)$\\
$\sqrt{e} \sin \omega$ & $\mathcal{U}(-1,1)$ & $\mathcal{U}(-1,1)$ & $\mathcal{U}(-1,1)$\\[1ex]
\multicolumn{4}{l}{\textbf{Jitter and Instrumental Terms:}} \\
$\log(\sigma_{{RV1}})$  & $\mathcal{N}(0,1)$ & $\mathcal{N}(0,1)$ & $\mathcal{N}(0,1)$ \\
$\log(\sigma_{RV2})$  & $\mathcal{N}(0, 1)$ & \nodata & \nodata\\
$\gamma_{\text{pre-HPF}}$ (km s$^{-1}$) & $\mathcal{N}(0,100)$ & $\mathcal{N}(0,100)$ & $\mathcal{N}(0,100)$ \\
$\gamma_{\text{post-HPF}}$ (km s$^{-1}$) & \nodata & $\mathcal{N}(0,100)$ & $\mathcal{N}(0,100)$ \\[1ex]
\hline
\multicolumn{4}{l}{\textbf{Derived Parameters}} \\
\hline
\multicolumn{4}{l}{\textbf{Orbital Parameters:}} \\
$P$ (days) & $126.948 \pm 0.029$ &  $9.686\pm0.001$  & $1876.750 \pm 23.634$ \\
$T_0$ (BJD$_{\text{TDB}}$) & $2459081.322 \pm 0.122$ & $2458739.118 \pm 0.321 $ & $2459662.762  \pm 0.945$ \\
$K_1$ (km s$^{-1}$) & $ 10.672 \pm 0.044$ & $ 0.167  \pm 0.013 $ &  $0.823  \pm 0.022$  \\
$K_2$ (km s$^{-1}$) & $13.551 \pm 0.226$ & \nodata & \nodata \\
$\omega$ (radians) & $-2.758\pm 0.006$ & $-0.246 \pm 0.242$ & $ 1.261  \pm 0.029$ \\
$\sqrt{e} \cos \omega$ & $-0.529 \pm 0.002$ & $ 0.243 \pm 0.069 $ & $0.208 \pm 0.018 $ \\
$\sqrt{e} \sin \omega$ & $-0.213 \pm 0.003$ & $ -0.061 \pm 0.055$ & $0.648 \pm  0.012$ \\
$e$ & $ 0.571 \pm 0.002 $ & $ 0.257 \pm 0.068$ & $0.681 \pm  0.009$ \\
\multicolumn{4}{l}{\textbf{Other Parameters:}} \\
$q$ & $ 0.788 \pm 0.013$ & \nodata & \nodata \\
$M_1\sin^3i (M_\odot)$  & $0.058 \pm 0.002$       & \nodata & \nodata\\
$M_2\sin^3i (M_\odot)$  &  $0.046  \pm 0.001$  & \nodata & \nodata\\
$a_1\sin i (AU)$ & $0.1022 \pm 0.0004$ & $0.000144 \pm 0.000009$ & $0.1040 \pm 0.0039$\\
$a_2\sin i (AU)$      & $0.1298 \pm 0.0021$      & \nodata & \nodata\\
$f (M_\odot)$      & $(4.0 \pm 0.38)\times10^{-5}$     & $(4.0 \pm 1.0)\times10^{-9}$ & $(8.85 \pm 0.12)\times10^{-3}$\\
\multicolumn{4}{l}{\textbf{Jitter and Instrumental Terms:}} \\
$\log(\sigma_{{RV1 (Pre-HPF)}})$  & $-2.820 \pm 0.351$ & $-3.657 \pm 0.400$ & $-3.897 \pm  0.383 $ \\
$\log(\sigma_{{RV1 (Post-HPF)}})$  & \nodata & $ -3.954 \pm 0.369$ & $-4.168   \pm0.378$ \\
$\log(\sigma_{{RV2}})$ (km s$^{-1}$) & $-1.266 \pm 0.561$ & \nodata & \nodata \\
$\gamma$$_{\text{pre-HPF}}$ (km s$^{-1}$) &  $-50.954\pm 0.021$ & $-11.996 \pm 0.013$  & $-32.689   \pm0.016$ \\
$\gamma$$_{\text{post-HPF}}$ (km s$^{-1}$) & \nodata & $-12.002\pm 0.008 $ & $-32.734  \pm 0.015$ \\
\multicolumn{4}{l}{\textbf{Primary Star Parameters$^b$:}} \\
$T_\mathrm{eff}$ (K) & $2875\pm59$ & $3137\pm59$ & $3322\pm59$ \\
$\log{g}$ & $5.17\pm0.04$ &  $5.01\pm0.04$ & $4.94\pm0.04$ \\
$[Fe/H]$ & $-0.02\pm0.16$ & $-0.06\pm0.16$ & $-0.15\pm0.16$ \\
Mass ($M_\odot$) & $0.104^{+0.006}_{-0.003}$ & $0.35^{+0.02}_{-0.07}$ & $0.32\pm{0.02}$ \\
Age (Gyr) & $9.3^{+2.9}_{-4.1}$ & $0.36^{+12.22}_{-0.36}$ & $2.68^{+10.04}_{-1.62}$
\enddata
\label{table:Stellar}
\tablenotetext{a}{Model priors follow $\mathcal{N}(\mu, \sigma)$ for normal and $\mathcal{U}(a, b)$ for uniform distributions.}
\tablenotetext{b}{The stellar parameters are derived with \texttt{HPF-SpecMatch} ($T_\mathrm{eff}$, $\log g$, $[Fe/H]$) or SED fitting and stellar evolution models (mass and age).}
\end{deluxetable*}

\subsection{LSPM J0515$+$5911}
\begin{figure}[t]
    \centering
    \includegraphics[width=1.0\linewidth]{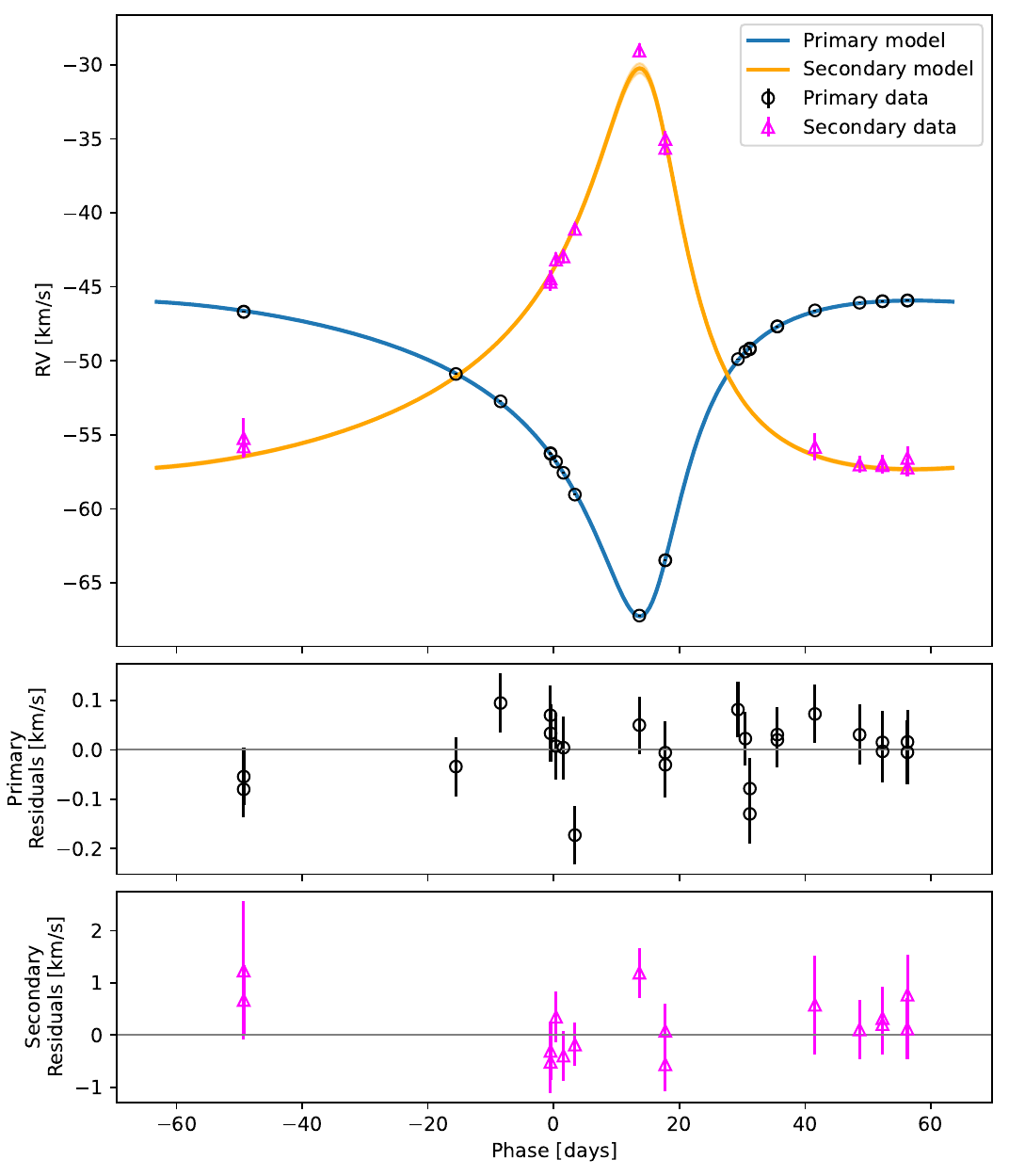}
    \caption{Radial velocity vs orbital phase for LSPM J0515+ 5911. Upper Panel: open black circles and open magenta triangles show our HPF primary and secondary velocities, respectively. The solid lines show the best orbital solution from Table \ref{table:Stellar}. Lower Panels: Observed –- calculated residuals for the  primary and secondary velocities.}
    \label{fig:LSPM}
\end{figure}

LSPM J0515$+$5911 consists of a primary M7 star in a binary system with a companion on a $126.948 \pm 0.029$ day eccentric orbit.  We solved this system as a double-lined spectroscopic binary. We used \texttt{HPF-Specmatch} to derive the stellar parameters based on a single high-S/N HPF observation that we obtained very close to the systemic velocity, and which yielded values of $T_{eff} = 2875 \pm 59 $K, $\log g = 5.17 \pm 0.04$, $[Fe/H] = -0.02 \pm 0.16 $, $v \sin i = 1 \pm 18~\mathrm{km\,s^{-1}}$. This corresponds to a primary template constructed from a spectrum of the star LSPM J0011$+$5908. We used observations obtained near quadrature to optimize the secondary template, by maximizing the correlation value, constructed from an HPF spectrum of the star GJ 644 C. We measured the primary velocity in 24 HPF spectra, and the secondary velocity in 16 of those spectra. The 8 spectra where we did not recover the secondary have low S/N or substantial blending between the two stars in velocity space.  Figure \ref{fig:LSPM} shows the primary and secondary RVs and corresponding orbital fits.
The eccentricity is  $e = 0.571 \pm 0.002$. The semi-amplitudes are $K_1 = 10.672 \pm 0.044$ km s$^{-1}$ and $K_2 = 13.551 \pm 0.226$ km s$^{-1}$, which gives the mass ratio of $q = M_2/M_1 = K_1/K_2 =  0.788 \pm 0.013$. The systemic velocity is $\gamma = -50.954 \pm 0.021$ km s$^{-1}$. All HPF spectra for this system were obtained in the `pre' era, so only a single systemic velocity is required to describe the measurements. 

The minimum masses from the RV orbit analysis are  $M_1\sin^3i =0.058 \pm 0.002$ $M_\odot$ and $M_2\sin^3i =0.046 \pm 0.001$ $M_\odot$. The posterior distributions for LSPM J0515$+$5911 are included in Figure \ref{fig:Corner_LSPM}. If these minimum masses were close to the true mass (i.e., if $i$ is close to 90$^\circ$), then both components would be sub-stellar brown dwarfs. This is implausible because the spectra of such objects would not be well represented by our HPF-Specmatch templates. Were the true mass of these objects in the L dwarf regime, the resulting correlation values would have been very small because the spectral lines would not be a good match. Using the SED fit procedure in \S3.2, we estimate the primary star's mass to be $0.104^{+0.006}_{-0.003}$ $M_{\odot}$ and age to be $9.3^{+2.9}_{-4.1}$ Gyrs. LSPM J0515$+$5911 must therefore be on an inclined orbit with an inclination of $i\sim48^\circ$ required to match the predicted stellar mass, which clearly excludes the possibility of eclipses.  This primary mass estimate, when combined with our measurement of $q$, gives a secondary mass of $M_2\sim 0.08\pm0.01\,M_\odot$, which is consistent with the temperature of stellar template we used for the cross-correlation analysis.

\subsection{NLTT 45468}\label{sec:nltt45468}

\begin{figure}[t]
    \centering
    \includegraphics[width=\linewidth]{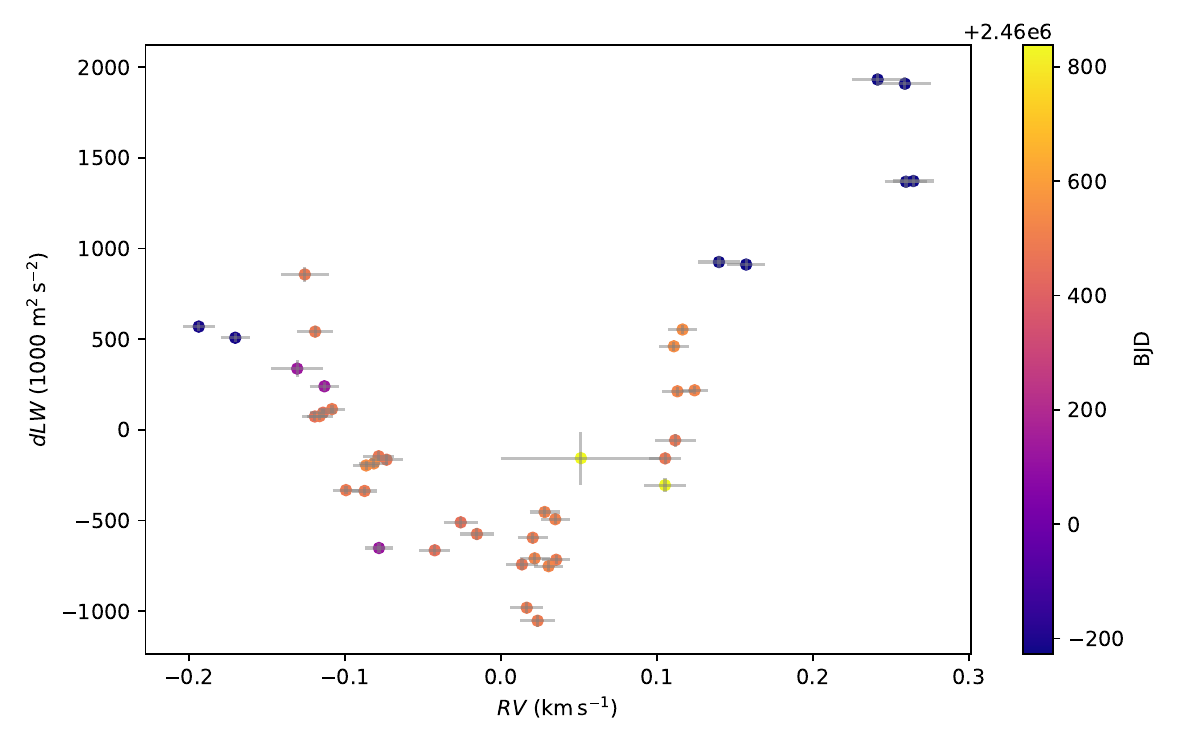}
    \caption{Differential line width ($\delta\mathrm{LW}$) vs. RV for NLTT 45468. The significant change in $\delta\mathrm{LW}$ suggests contamination in the spectra, possibly from the nearby star at 2".}
    \label{fig:linewidth}
\end{figure}

NLTT 45468 is an M4 type star in a binary system with a period of $9.686\pm0.001$ days. The nearby star shown in Figure~\ref{fig:threeplots}b at a distance of $\sim$2\arcsec{} is close enough to cause some contamination in the 1.7\arcsec{} HPF fiber, but the effect is difficult to quantify when combined with variable seeing and guiding over years of observations. This object is resolved in Gaia (Gaia DR3 1435275918821040000) with a parallax of 34.56 mas, nearly identical to NLTT 45468, and a $\Delta\mathrm{G_{mag}}\sim2$, which suggests that this is likely a bound wide companion, with a velocity close to the systemic velocity of the short period system. To assess the potential impact on our RV measurements, we analyzed the HPF spectra of NLTT 45468 using HPF-SERVAL \citep{2020AJ....159..100S,2023Sci...382.1031S}, which provides measurements of the differential line width, $\delta\mathrm{LW}$, which is the second moment of the cross-correlation function \citep{2018A&A...609A..12Z}. Variations in $\delta\mathrm{LW}$ are attributable to variations in the shape of the spectral lines, often due to stellar activity, but also possibly due to a nearby star whose contamination level varies between observations. Figure~\ref{fig:linewidth} shows a strong correlation between $\delta\mathrm{LW}$ and the RVs, suggesting that there is indeed contamination from the companion affecting the correlations. This contamination will manifest as a weak correlation signal at a velocity near the primary overlapping which would reduce the signal to noise of the secondary and potentially cause a bias in the velocity measured of the secondary star if we were able to do so.

We solved NLTT 45468 as a single lined spectroscopic binary; no secondary component was evident in our \texttt{SXCORR} analysis.  We used \texttt{HPF-Specmatch} to derive the stellar parameters to be $T_{eff} = 3127 \pm 59 $K, $\log g = 5.03 \pm 0.04$, and $[M/H] = -0.05 \pm 0.16 $, $v \sin i = 4 \pm 1~\mathrm{km\,s^{-1}}$, and used a template constructed from an HPF spectrum of GJ 9066. The eccentricity of the SB1 orbit is  $e = 0.2571 \pm 0.068$. The semi-amplitude is $K_1 = 0.167 \pm 0.013$ km s$^{-1}$. The systemic velocities for the pre and post HPF eras are $\gamma_{preHPF} = -11.996 \pm 0.013$ km s$^{-1}$ and $\gamma_{postHPF} = -12.002 \pm 0.008$ km s$^{-1}$, respectively. The minimum value of the semi-major axis of the primary star's orbit is $a_1\sin i = 0.031 \pm 0.002 $ $R_{\odot}$. The posterior distributions for NLTT 45468 are included in Appendix \ref{sec:appendix}, Figure \ref{fig:Corner_NLTT45468}.
  From the SED fit, we estimate the primary star's mass to be $0.35^{+0.02}_{-0.06}$ $M_{\odot}$ and age to be $0.36^{+12.22}_{-0.36}$ Gyrs.

\begin{figure}[t]
    \centering
    \includegraphics[width=1.0\linewidth]{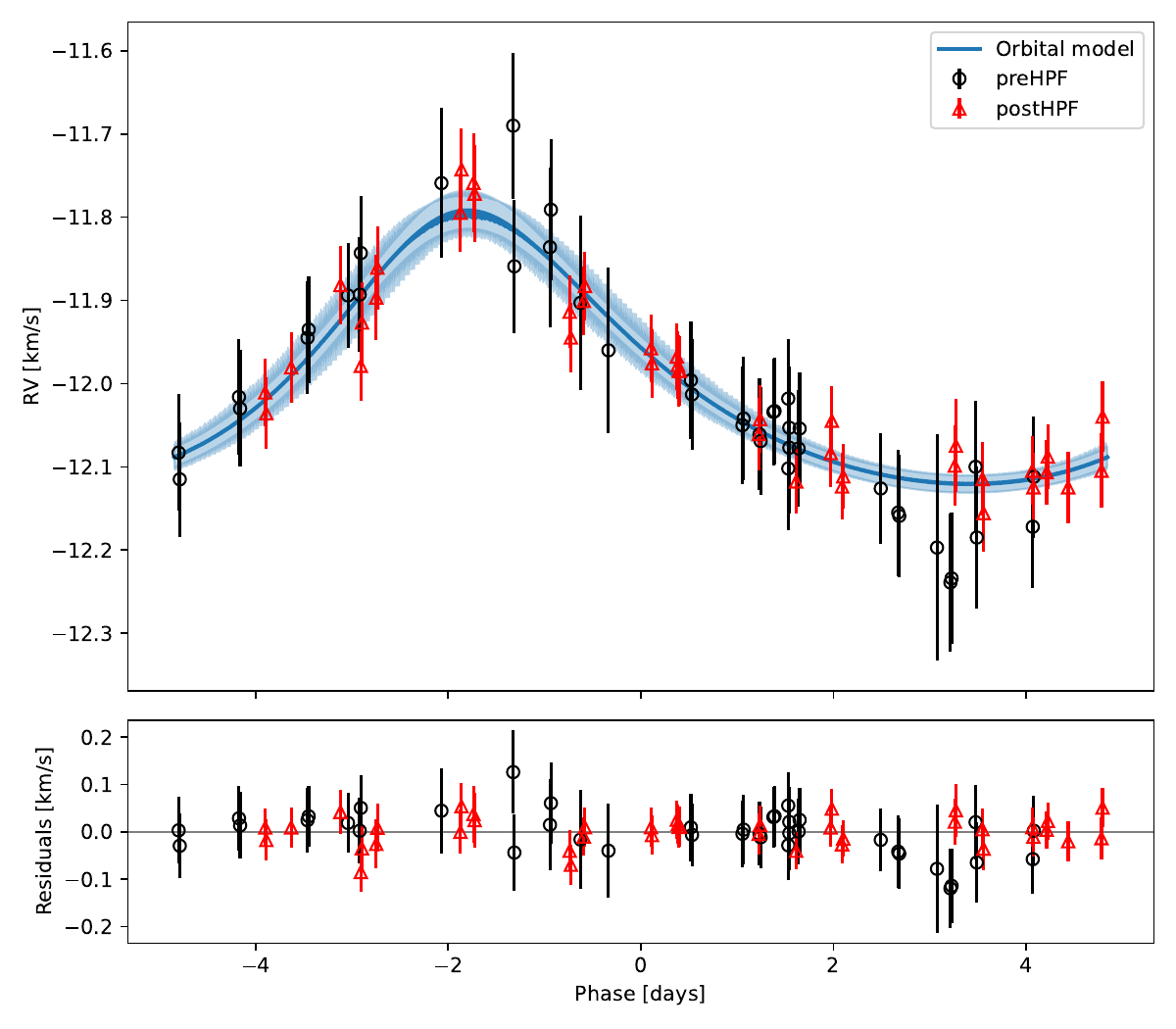}
    \caption{RV vs orbital phase for NLTT 45468. \textit{Upper Panel:}  Open black circles and open red triangles show primary HPF $V_{pre}$ and $V_{post}$, respectively, corrected for the relative offset to be on the plotted on the same plot.
The solid lines show the best orbital solution (Table \ref{table:Stellar}).
\textit{Lower Panels:} Observed - calculated residuals for the HPF pre and post primary RVs}
    \label{fig:RV-NLTT45468}
\end{figure}

The relatively short 9.7 day orbital period of NLTT 45468 opens the possibility for a geometric alignment that would cause eclipses. We  analyzed the light curves from TESS-SPOC with 41 sectors of data, adapting the procedure used in \cite{2023Sci...382.1031S}. Our analysis included processing of data by doing a sigma clipping, with upper bound = 3 and lower bound = 5. We then detrended the light curves using Wotan\footnote{\url{https://github.com/hippke/wotan}} \citep{2019AJ....158..143H} keeping the window length = 0.50 and then performed a transit least squares (TLS)\footnote{\url{https://github.com/hippke/tls}} search \citep{2019A&A...623A..39H}. We adopted the recommended signal detection efficiency threshold of 7 (corresponding to a 1\% false positive rate) with TLS. Due to the difference in noise properties across sectors, we performed this analysis for each individual sector of data. We did not find any significant and consistent transit events in any of the 41 sectors.

This low-mass binary presents an interesting opportunity for followup observations at higher angular and spectral resolution. An adaptive optics fed spectrograph optimized for the near-infrared, such as iLocater \citep{2022SPIE12184E..1PC} on the Large Binocular Telescope, could easily eliminate the signal from the wide companion that contaminates HPF spectra, and at the same time apply higher-spectral resolution to reduce blending in velocity space between the short period objects.  Alternatively, a seeing limited spectrograph with a smaller diameter fiber, such as NEID on WIYN telescope, with R$\sim$115,000, may also be able to detect a signal from the secondary.

\subsection{NLTT 43564\label{sec:nltt43564}}

\begin{figure}[!t]
    \centering
    \includegraphics[width=1.0\linewidth]{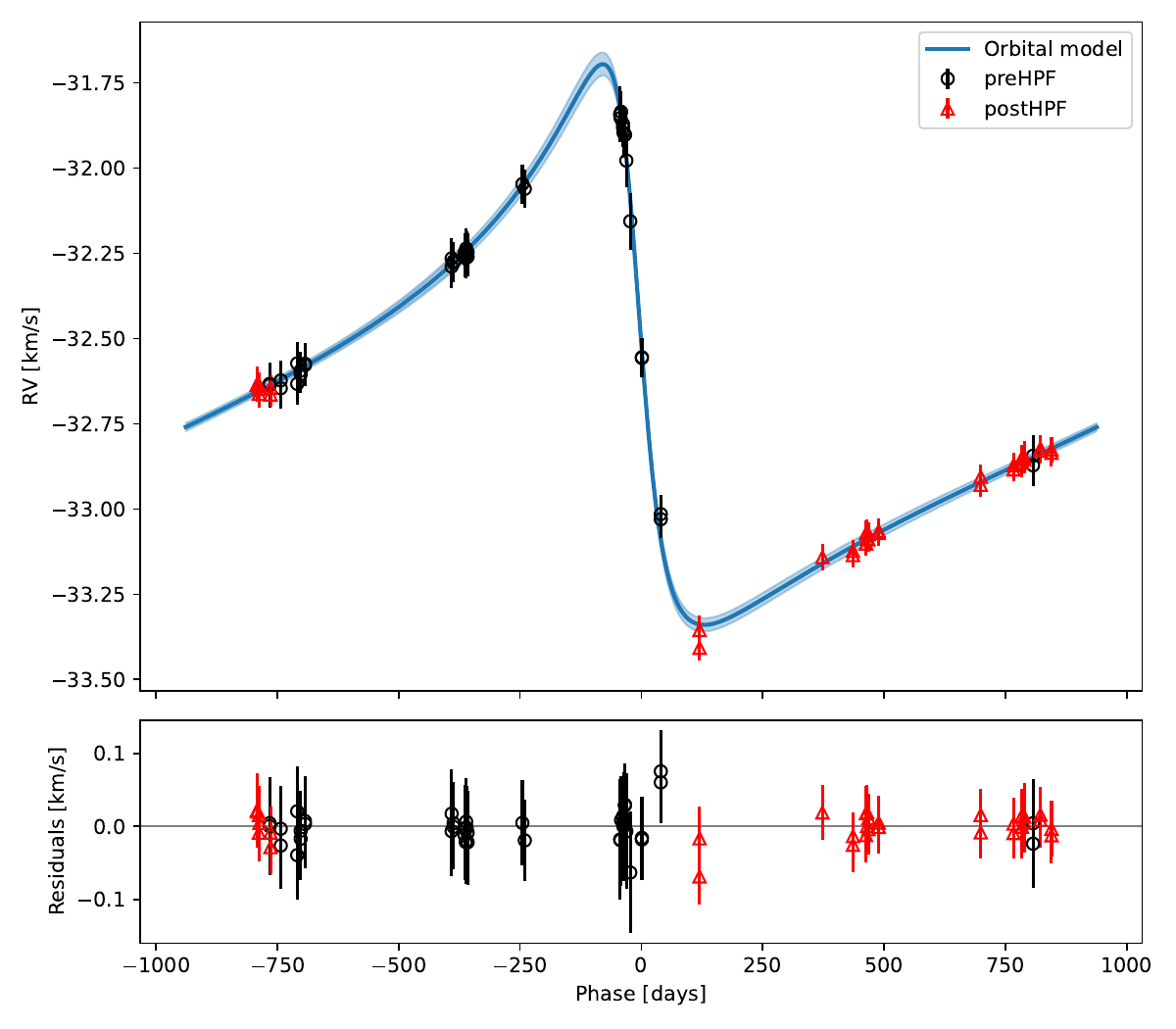}
    \caption{Same as Figure~\ref{fig:RV-NLTT45468}, but for NLTT 43564.}
    \label{fig:RV-43564}
\end{figure}

NLTT 43564 is an M4 type star in a highly eccentric binary system with a period of $1870 \pm 24$ days. We solved NLTT 43564 as a single-lined spectroscopic binary; no secondary component was evident in our \texttt{SXCorr} analysis. We used \texttt{HPF-Specmatch} to derive the stellar parameters of the primary star to be: $T_{eff} = 3322 \pm 59 $K, $\log g = 4.94 \pm 0.04$, and $[M/H] = -0.15 \pm 0.16 $, $v \sin i = 1.4 \pm 0.4~\mathrm{km\,s^{-1}}$ and used a template constructed from an HPF spectrum of GJ 105.  Figure \ref{fig:RV-43564} shows the phased velocities and our derived orbital solution. Because of the very long orbital period, the phased plot is nearly identical to an unphased plot: the first observation occurs around +800 days, and the last observation occurs near -700 days, corresponding to a total span of just over one full orbit.

This system is highly eccentric: $e = 0.681 \pm 0.009$. The semi-amplitude is $K_1 = 0.823 \pm 0.022$ km s$^{-1}$. The systemic velocities pre and post HPF era are $\gamma_{preHPF} = -32.689 \pm 0.016$ km s$^{-1}$ and $\gamma_{postHPF} = -32.734 \pm 0.015$ km s$^{-1}$, consistent to within the 1$\sigma$ uncertainties. 
From the SED fit analysis, we estimate the primary star's mass to be $0.32 \pm 0.02$ $M_{\odot}$ and age to be $2.68^{+10.04}_{-1.62}$ Gyrs.

This system has a long orbital period, which opens the possibility of detecting astrometric motion that could be combined with RVs to derive a fully dynamical orbital solution. Gaia DR3 reports a large RUWE of 17.164, which is consistent with this system being a binary \citep{2018A&A...616A...2L}. Gaia DR4, which is anticipated in late 2026, should significantly improve on the astrometric precision achieved for binaries like NLTT 43564, and so could enable a more model independent analysis of the system's orbital parameters. However, to achieve a fully dynamical solution would still require detecting the companion spectroscopically. Such a measurement could be attempted by a targeted campaign with HPF or other RV spectrograph for very high-S/N spectra near binary quadrature.

\subsection{Additional analyses}

\begin{figure}[!t]
    \centering
    \includegraphics[width=1\linewidth]{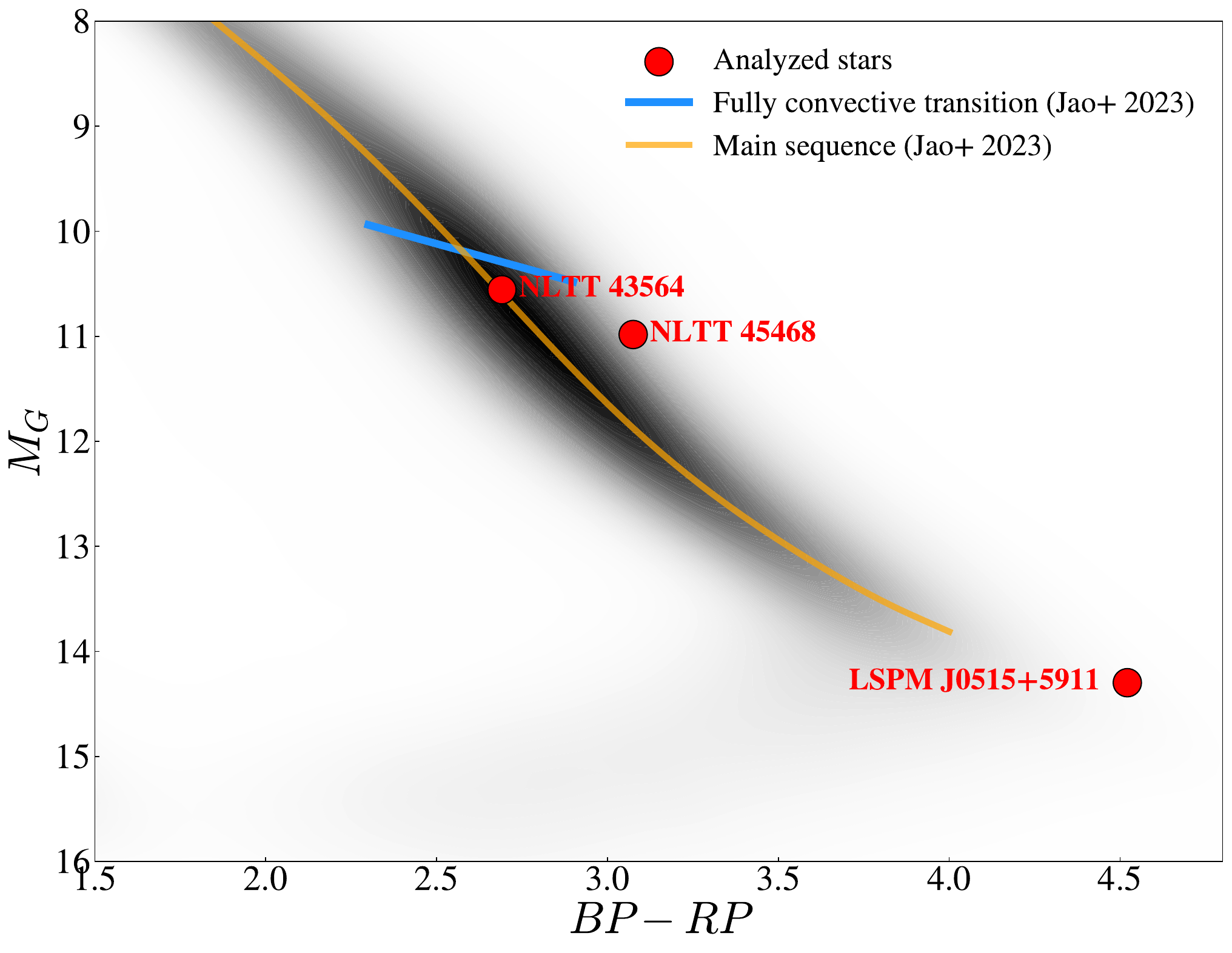}
    \caption{
Color-magnitude diagram for nearby stars (within 100 pc) showing the local main sequence in $M_G$ vs $BP-RP$. The background density map is constructed from the Gaia Catalogue of Nearby Stars \citep{2021A&A...649A...6G}. The three analyzed binary star targets, LSPM~J0515+5911, NLTT~45468, and NLTT~43564, are plotted as red markers and they include light from the unresolved companion as seen by Gaia. The  blue line indicates the observational transition to fully convective M dwarfs and the orange curve shows the empirical main-sequence points from \citet{Jao}. 
}
    \label{fig:hrdiagram}
\end{figure}

To better understand the properties of the three binary systems we have presented here in the context of other solar neighborhood systems, we have plotted them in Figure \ref{fig:hrdiagram} their blended light colors as seen by Gaia on a color-magnitude diagram generated from Gaia DR3 measurements of stars within 100 pc of the Sun \citep{2021A&A...649A...6G}. We also show the positions of the main sequence and the transition between partially and fully convective stellar interiors as parameterized by \citep{Jao}. All of our stars are cool, fully-convective, M dwarfs. NLTT 43564 lies close to the HR gap, which provides further motivation for future efforts to solve this system fully dynamically. LSPM J0515+5911 sits at a very interesting position at the bottom of the main-sequence, but its moderate inclination, medium period, and relative faintness combine to make followup observations that could solve this system fully dynamically quite challenging.  

For single-lined binaries where we are unable to detect a signal from the secondary component in our spectra, we can estimate an upper limit on the secondary mass by considering the excess astrometric noise, $\epsilon$, measured by Gaia, attributed to an offset between the center of light and the center of mass due to Keplerian orbital motion \citep{2020MNRAS.495..321P}.  For NLTT 45468, the simplification where the orbital period is much less than the observing window  of Gaia \citet{2020MNRAS.495..321P} Eq.~17 applies. In the limit where the secondary is ``dark'', which is consistent with our lack of detection in the HPF spectra, we can set the luminosity ratio $l=\mathrm{L_2/L_1}=0$. Following the parameterization in \citep[][Eq.~2]{2022AJ....163...89C}, with $\epsilon\sim0.09208\,\mathrm{mas}$ and the orbital parameters in Table~\ref{tab:priors_results}, we estimate an upper limit on the secondary mass for NLTT 45468 of $M_2=0.016\,M_\odot$, suggesting that the companion is substellar. If we consider an upper limit of $l=0.05$, which would easily be detectable in our HPF spectra, the upper limit is $M_2=0.035\,M_\odot$, which is still quite securely sub-stellar. For NLTT 43564, the long orbital period means that the full solution for long period binaries in \citet[][Appendix B]{2020MNRAS.495..321P} must be used. We used the Gaia Observation Forecast Tool\footnote{https://gaia.esac.esa.int/gost/index.jsp} to estimate the dates that Gaia observed NLTT 43564 during the time period represented by DR3. With $\epsilon=2.822\,\mathrm{mas}$, and making the same $l=0.0$ limit as above, we estimate an upper limit on $M_2=0.022\,M_\odot$. In the upper limit of $l=0.05$, we estimate $M_2=0.035\,M_\odot$.  For both of these binaries, the Gaia excess astrometric noise suggests that the secondary objects are substellar. This further motivates the future work suggested in \S\ref{sec:nltt45468} and \S\ref{sec:nltt43564} to push for additional dynamical constraints on these systems.

The age in all of our M dwarf targets are weakly constrained. However, the rotational properties of low-mass stars provide a useful diagnostic. As M dwarfs evolve, they gradually lose angular momentum through magnetic braking, causing their rotation periods to increase over time \citep{2011ASPC..451..285E}. This spin-down behavior enables age estimation via gyrochronology \citep{2003ApJ...586..464B, 2007AcA....57..149K, 2016ApJ...821...81G, 2021ApJ...916...77P}. For early-type M dwarfs (M0--M1~V), \citet{2023ApJ...954L..50E} derived an empirical relation connecting rotation period to stellar age. However, in the mass regime for the stars presented here, this relationship is complicated \citep[e.g.,][]{2016csss.confE...4N,2023ApJ...954L..50E,2023MNRAS.520.5283G}. All three systems lie in the regime in the color-magnitude diagrams in \citet{2021ApJ...916...77P} of $G$ vs $G-G_{RP}$ that correspond to field star populations with unmeasured rotational velocities. Reliably measuring these from our HPF spectra at R$\sim$53,000 would be extremely difficult.  This is consistent with the ages we report in Table~\ref{tab:priors_results}, which are extremely uncertain, but consistent with main-sequence stars.

\section{Summary\label{sec:summary}}
We used ground-based observations from HPF, NESSI, 'Alopeke, and Robo-AO, to derive the orbital parameters and constrain the stellar components of three spectroscopic binary systems comprised of fully-convective stars: LSPM J0515+5911, NLTT 45468, and NLTT 43564. LSPM J0515+5911 and NLTT 43564 are particularly interesting because their orbital periods are sufficiently long to suggest that the individual components are not tidally interacting, and so evolved as single stars. We solved LSPM J0515+5911 as an SB2 and determined that it is comprised of two late-type M dwarf stars in an eccentric orbit, with masses of $M_1=0.104\pm0.01\,M_\odot$ and $M_2=0.08\pm0.01\,M_\odot$. We solved NLTT 43564 and NLTT 45468 as SB1s, and estimate the primary masses to be $M_1\sim0.35\,M_\odot$ and $M_1\sim0.32\,M_\odot$, respectively. Each of these masses are consistent with the stellar temperatures that we derived from our HPF spectral template analyses.  NLTT 45468 has a nearby companion that contaminates our seeing-limited HPF spectra, making it an ideal candidate for followup observations with an AO spectrograph such as iLocater. NLTT 43564 has a period of more than 5 years, and is a promising candidate for a full astrometric + spectroscopic analysis upon the release of Gaia DR4. 

These binaries are part of a large observational program using HPF to followup eclipsing and non-eclipsing low-mass binary stars. Most of these targets were originally included in programs searching for exoplanets around M dwarfs. We have continued observing them, even after the signature of a stellar companion was detected, because of the utility of dynamical measurements of low-mass stellar binaries. Some targets that we continue to monitor, and plan to publish in future papers, have periods even longer than NLTT 43564, and so are prime candidates for combining results from spectroscopy and astrometry to yield fully dynamical solutions.

\begin{acknowledgments}
CIC acknowledges support from NASA Headquarters (i) through an appointment to the NASA Postdoctoral Program at the Goddard Space Flight Center, administered by ORAU through a contract with NASA and (ii) under award number 80GSFC24M0006.

Resources supporting this work were provided by the Pennsylvania State University's Institute for Computational and Data Sciences' (ICDS) Roar supercomputer. This content is solely the responsibility of the authors and does not represent the views of ICDS.

We acknowledge support from NSF grants AST 1006676, AST 1126413, AST 1310875, AST 1310885, AST 2009554, AST 2009889, AST 2108512, AST 2108801 and the NASA Astrobiology Institute (NNA09DA76A) in our pursuit of precision RVs in the near-infrared. We acknowledge the support of the Heising-Simons Foundation through grant 2017-0494. These results are based on observations obtained with HPF on the HET. The HET is a joint project of the University of Texas at Austin, the Pennsylvania State University, Ludwig-Maximilians-Universit\"at M\"unchen, and Georg-August Universit\"at Gottingen. The HET is named in honor of its principal benefactors, William P. Hobby and Robert E. Eberly. The HET collaboration acknowledges the support and resources from the Texas Advanced Computing Center. We are grateful to the HET Resident Astronomers and Telescope Operators for their valuable assistance in gathering our HPF data.
We would like to acknowledge that the HET is built on Indigenous land. Moreover, we would like to acknowledge and pay our respects to the Carrizo \& Comecrudo, Coahuiltecan, Caddo, Tonkawa, Comanche, Lipan Apache, Alabama-Coushatta, Kickapoo, Tigua Pueblo, and all the American Indian and Indigenous Peoples and communities who have been or have become a part of these lands and territories in Texas, here on Turtle Island.

Some of the observations were obtained at the WIYN Observatory from telescope time allocated to NN-EXPLORE (PI: Gupta; 2022A-665981) through the scientific partnership of NASA, the NSF, and NOIRLab. These observations used the NN-EXPLORE Exoplanet and Stellar Speckle Imager (NESSI). NESSI was funded by the NASA Exoplanet Exploration Program and the NASA Ames Research Center. NESSI was built at the Ames Research Center by Steve B. Howell, Nic Scott, Elliott P. Horch, and Emmett Quigley.
WIYN is a joint facility of the University of Wisconsin-Madison, Indiana University, NSF's NOIRLab, the Pennsylvania State University, Purdue University, the University of California-Irvine, and the University of Missouri. The authors are honored to be permitted to conduct astronomical research on Iolkam Du'ag (Kitt Peak), a mountain with particular significance to the Tohono O'odham.
\end{acknowledgments}

Some of the data presented in this paper were obtained from MAST at STScI. Support for MAST for non-HST data is provided by the NASA Office of Space Science via grant NNX09AF08G and by other grants and contracts.
This work includes data collected by the TESS mission, which are publicly available from MAST. Funding for the TESS mission is provided by the NASA Science Mission directorate. 
This research made use of the (i) NASA Exoplanet Archive, which is operated by Caltech, under contract with NASA under the Exoplanet Exploration Program, (ii) SIMBAD database, operated at CDS, Strasbourg, France, (iii) NASA's Astrophysics Data System Bibliographic Services, (iv) NASA/IPAC Infrared Science Archive, which is funded by NASA and operated by the California Institute of Technology, and (v) data from 2MASS, a joint project of the University of Massachusetts and IPAC at Caltech, funded by NASA and the NSF.

This work has used data from the European Space Agency (ESA) mission Gaia (\url{https://www.cosmos.esa.int/Gaia}), processed by the Gaia Data Processing and Analysis Consortium (DPAC, \url{https://www.cosmos.esa.int/web/Gaia/dpac/consortium}). Funding for the DPAC has been provided by national institutions, in particular the institutions participating in the Gaia Multilateral Agreement.

\vspace{5mm}
\facilities{TESS, RBO, WIYN (NESSI), HET (HPF), Gemini:Gillett ('Alopeke), Robo-AO, MAST}

\software{
Astropy              \citep{astropy1, astropy2, 2022ApJ...935..167A},
HPF-SERVAL           \citep{gudmundur_stefansson_2023_8397170},
lightkurve           \citep{2018lightkurve},
Matplotlib           \citep{matplotlib},
NumPy                \citep{numpy},
pandas               \citep{pandas},
SciPy                \citep{scipy1, scipy2},
HPF-Specmatch         \citep{Stefansson_2020},
PyMC3                \citep{2015arXiv150708050S},
exoplanet            \citep{2021JOSS....6.3285F},
astroARIADNE        \citep{2022MNRAS.tmp..920V},
tls                 \citep{2019A&A...623A..39H}
}

\appendix
\section{Additional Analysis}
\label{sec:appendix}
These corner plots summarize the posterior distributions obtained from our exoplanet modeling described above. We include them for completeness here. 
\begin{figure*}
    \centering
    \includegraphics[width=1\linewidth]{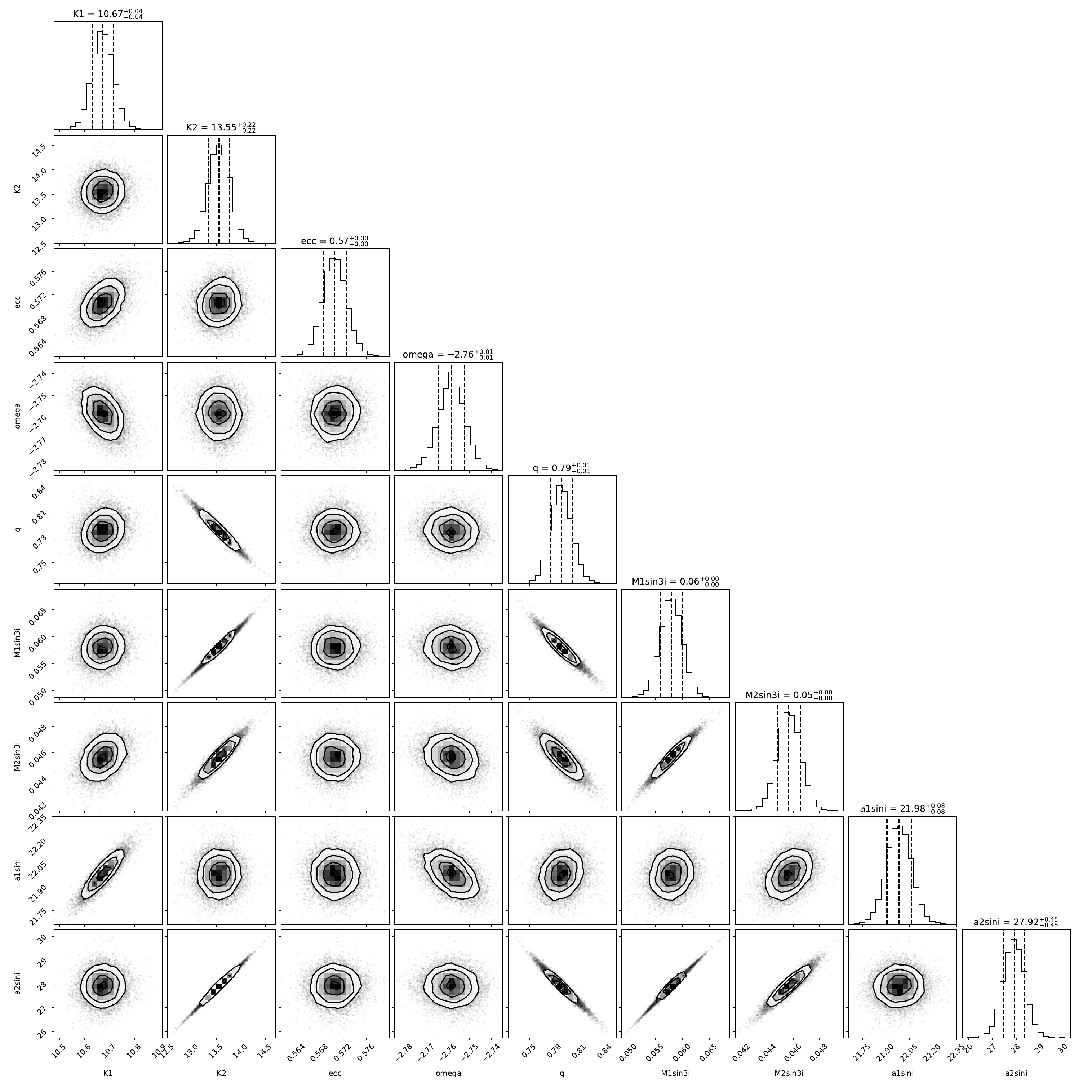}
    \caption{Posterior distributions obtained for LSPM J0515$+$5911, which was solved as a double-lined binary.}
    \label{fig:Corner_LSPM}
\end{figure*}
\begin{figure*}
    \centering
    \includegraphics[width=1\linewidth]{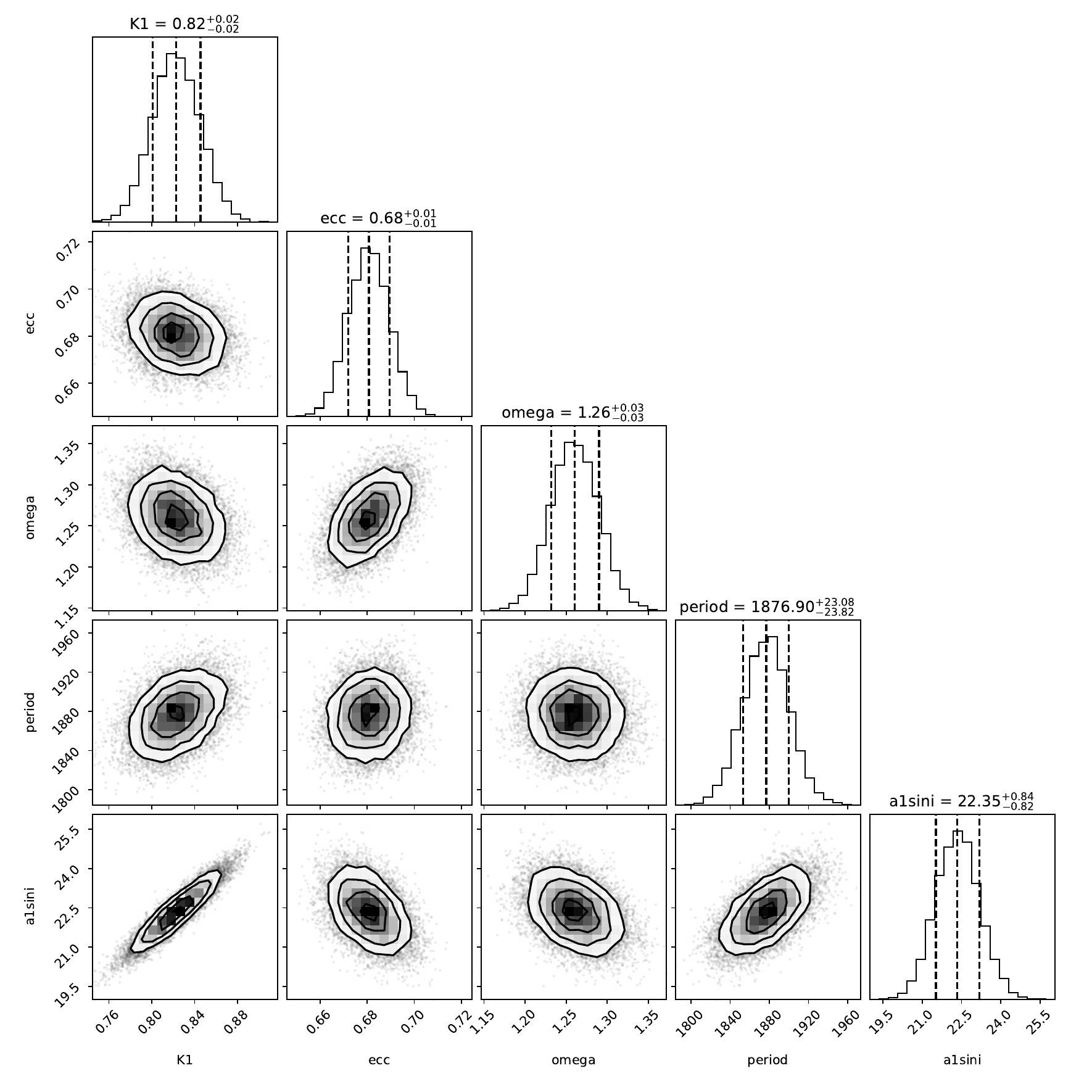}
    \caption{Posterior distributions obtained for NLTT 43564, which was solved as a single lined binary.}
    \label{fig:Corner_NLTT43564}
\end{figure*}
\begin{figure*}
    \centering
    \includegraphics[width=1\linewidth]{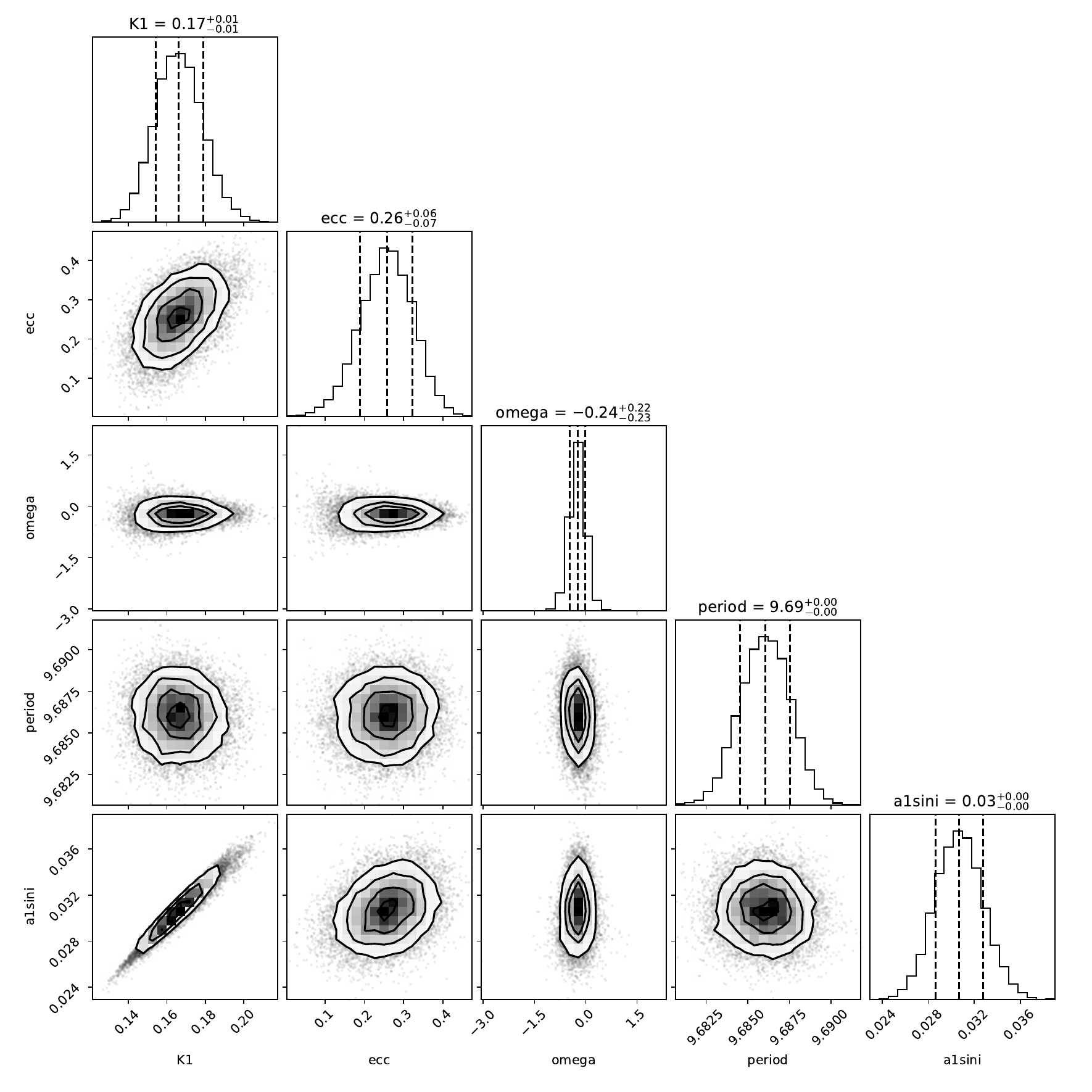}
    \caption{Posterior distributions obtained for NLTT 45468, which was solved as a single lined binary.}
    \label{fig:Corner_NLTT45468}
\end{figure*}

\bibliography{main}{}
\bibliographystyle{aasjournal}

\end{document}